\newcommand{\be}{\begin{equation}}
\newcommand{\ee}{\end{equation}}
\newcommand{\bea}{\begin{eqnarray}}
\newcommand{\eea}{\end{eqnarray}}
\newcommand{\ba}{\begin{array}}
\newcommand{\ea}{\end{array}}

\newcommand{\al}{\alpha}
\newcommand{\ga}{\gamma}
\newcommand{\Ga}{\Gamma}

\newcommand{\ka}{\kappa}
\newcommand{\de}{\delta}
\newcommand{\vphi}{\varphi}
\newcommand{\ep}{\epsilon}
\newcommand{\si}{\sigma}
\newcommand{\la}{\lambda}

\newcommand{\Om}{\Omega}
\newcommand{\ze}{\zeta}

\newcommand{\La}{\Lambda}

\newcommand{\tha}{\theta}

\newcommand{\Ups}{\Upsilon}

\newcommand{\tphi}{\tilde{\phi}}
\newcommand{\tX}{\tilde{X}}

\newcommand{\tr}{{\rm tr}}
\newcommand{\Z}{\mathbb{Z}}
\newcommand{\D}{{\rm d}}
\newcommand{\pa}{\partial}
\newcommand{\rar}{\rightarrow}
\newcommand{\non}{\nonumber}
\newcommand{\we}{\kern-.1em\wedge\kern-.1em}
\newcommand{\Li}{\mathrm{Li}}
\newcommand{\cO}{\mathcal{O}}

\newcommand{\half}{\mbox{$\frac{1}{2}$}}
\newcommand{\B}{|B\rangle}
\newcommand{\BF}{|B(F)\rangle}
\newcommand{\ts}{\textstyle}
\newcommand{\ds}{\displaystyle}
\newcommand{\II}{I\kern-.09em I}

\documentclass[12pt]{article}

\usepackage{amsfonts}
\usepackage{pst-node}
\usepackage{cite}

\begin{document}
 
\begin{flushright}
 DAMTP-2000-80 \\ {\tt hep-th/0008125}
\end{flushright}
\vspace{1mm}
\begin{center}{\bf\Large\sf Derivative corrections to D-brane actions
\\ with constant background fields}
\end{center}
\vskip 5mm
\begin{center}
Niclas Wyllard\footnote{\tt N.Wyllard@damtp.cam.ac.uk} \vspace{5mm}\\
{\em  Department of Applied Mathematics and Theoretical Physics, \\ Centre for Mathematical Sciences, University of
Cambridge, \\ Wilberforce Road, Cambridge, CB3 0WA, United Kingdom}
\end{center}
 
\vskip 5mm
 
\begin{abstract}
We study derivative corrections to the effective action for a single D-brane
in type~\II{} superstring theory coupled to constant background
fields. In particular, within this setting we determine the complete
expression for the (disk-level) four-derivative corrections to the
Born-Infeld part of the action. We also determine $2n$-form
$2n$-derivative corrections to the Wess-Zumino term. Both types of
corrections involve all orders of the gauge field strength, $F$. The
results are obtained via string $\si$-model loop calculations using
the boundary state operator language. The corrections can be
succinctly written in terms of the Riemann tensor for a non-symmetric
metric. 

\end{abstract}

\setcounter{equation}{0}
\section{Introduction}

Most work on D-brane effective actions have focused on the lowest
order terms in a derivative expansion. For the case of a single
brane in type \II{} superstring theory, these terms have been
completely determined and are well known. They are of two types: the
parity-violating Wess-Zumino term \cite{Douglas:1995}, which includes couplings to the
Ramond-Ramond background fields, and the parity-conserving
Born-Infeld part
\cite{Fradkin:1985,Abouelsaood:1987,Leigh:1989}. The
inclusion of fermions to form a  $\ka$-symmetric
action (including the couplings to a general
supergravity background) is known
\cite{Cederwall:1996b}. It has
also been shown (for the lowest order terms) that T-duality is realised in the D-brane effective
action \cite{Alvarez:1996,Green:1996}. The lowest order action is in fact highly constrained and
can be  completely determined in a variety of ways. Apart from explicit
calculations
\cite{Fradkin:1985,Abouelsaood:1987,Leigh:1989,Garousi:1998}, these
terms can also be  fixed completely using the constraints of supersymmetry in the guise of
$\ka$-symmetry \cite{Cederwall:1998}, and also (for the Born-Infeld part) by Lorentz
covariance combined with T-duality \cite{Bachas:1996} or  via
the connection to its non-commutative version \cite{Seiberg:1999}. The
requirement of electromagnetic Montonen-Olive duality (or more
generally $\mathrm{SL}(2,\mathbb{Z})$ invariance) of the D3-brane also
puts strong constraints on the form of the action \cite{Gibbons:1995,Tseytlin:1996}. Together with certain conditions on the solutions it uniquely leads to the
Born-Infeld action \cite{Deser:1998}.  In addition, the Wess-Zumino part can be
determined by anomaly cancellation arguments
\cite{Green:1997}. 

There are in general derivative
corrections to the lowest order terms, which, because of the
fundamental role played by the D-branes (see
\cite{Polchinski:1996} for reviews), are presumably important and
may possibly lead to a deeper understanding of the properties of D-branes and
various dualities.
The derivative corrections fall into two classes. Firstly, there are corrections involving derivatives of the
world-volume fields, i.e the gauge field $F_{\mu\nu}$ and the
embedding fields $\pa_{\mu} X^{i}$, where $X^i$ are the transverse scalars\footnote{We will throughout this
paper consider $F_{\mu\nu}$ and $\pa_{\mu}X^{i}$ to be of zeroth
order. When we speak of derivative corrections we
thus mean corrections which vanish when $F$ and $\pa X$ are
constant.}. Secondly, there are corrections which involve the pullback
of expressions containing derivatives of the bulk fields ($g_{MN}$,
$B_{MN}$, $\Phi$ and the $C_{p}$'s) to the
world volume of the brane. In general there are of course also combinations of the two types.
Corrections of the second type were discussed in \cite{Bachas:1999}, where
$\al'^2$ corrections\footnote{We absorb a
factor of $2\pi\al'$ into $F$ and $\pa X$ to make them
dimensionless; $\al'^2$ corrections thus involve four derivatives.} to the Born-Infeld part arising from the
gravity sector of the superstring background were determined by comparison with
earlier string calculations \cite{Garousi:1996,Garousi:1998}. These corrections involve quadratic expressions of the pullbacks of the
bulk Riemann tensor to the normal and tangent bundles. Corrections to
the Wess-Zumino term involving the bulk Riemann tensor were determined
using anomaly cancellation arguments in
\cite{Green:1997} and have subsequently also been obtained via direct string
theory calculations \cite{Craps:1998}. In \cite{Bachas:1999} corrections
involving the second fundamental form of the embedding were also
determined (both for the Wess-Zumino and Born-Infeld parts of the
action). Since the second fundamental form is constructed out of derivatives of
the embedding fields, certain corrections of the first type were thus
obtained. 

In this paper we study derivative corrections to the gauge sector of the
effective action of a single D-brane in type \II{} superstring theory. (We
only consider branes which preserve half of the supersymmetries and
not the recently much studied ``non-BPS'' D-branes. We also do not
consider the non-abelian case.) There are several different
methods available for calculating such corrections. Firstly, there is
the string S-matrix method \cite{Neveu:1972}, where one proceeds by directly calculating string
scattering amplitudes and from the result reconstruct the effective
action. 
 String S-matrix techniques are powerful, but for the D-branes they have the
drawback that they are perturbative in powers of the gauge field
strength $F$; fortunately, there are other methods available which are non-perturbative in powers of the gauge field $F$,
i.e. they are powerful enough to allow one to sum  the entire
perturbative series. One of these methods is the so called $\beta$-function method, where one calculates the $\beta$-function of the string $\si$-model as a loop expansion and interpret
its vanishing as equations of motion for the world-volume fields. This approach is closely related
to the requirement of  conformal invariance. 
Using this method the Born-Infeld part of the lowest order action  was
obtained for the ten-dimensional case (or in modern language, for the
D9-brane) in \cite{Abouelsaood:1987} and for the lower-dimensional
D-branes in \cite{Leigh:1989}. A method in the same vein is the
partition function method
\cite{Fradkin:1985,Tseytlin:1986b,Andreev:1988}. This was the method
used when the parity-conserving part of the 
effective action was first obtained for the nine-brane case in
\cite{Fradkin:1985} and has subsequently been extended to the lower-dimensional cases as
well \cite{Tseytlin:1996}. For both the superstring and
the bosonic string this method  (for the lowest order
term, i.e. neglecting derivatives) leads to the well known Born-Infeld
determinant form. When derivative corrections are included the
method gives the wrong result for the case of  the bosonic
string \cite{Andreev:1988}. For the superstring case, however, it has
been shown that the
method gives correct results \cite{Andreev:1988}. Closely related to the partition function method is the boundary state formalism
\cite{Callan:1987a,Callan:1988,Green:1994} (see \cite{DiVecchia:1999a} for a recent review). The
boundary state is an important tool in the study of D-branes in type
\II{} superstring theory and
encodes various properties of the D-branes, including the effective
action. The boundary state formalism has the advantage that it leads
to a natural way of determining the Wess-Zumino term, viz. by calculating the
overlap of the state representing the RR form
fields with the boundary state.  The extension of the boundary state method to incorporate a non-constant gauge field was recently discussed in detail \cite{Hashimoto:1999a,Hashimoto:1999b} (see also \cite{Callan:1988}).

For the D-branes in type \II{} superstring theory the $\cO(\al')$
corrections involving the gauge field $F$ to the Born-Infeld part of the
action are known to vanish \cite{Andreev:1988,Hashimoto:1999b}. The
$\cO(\al')$ corrections to the Wess-Zumino term also vanish (see e.g.
\cite{Hashimoto:1999b}). Thus, the first non-trivial corrections appear
at order $\al'^2$. The $\al'^2$ corrections to the Born-Infeld term
involving four
$F$'s (and four derivatives) are known \cite{Andreev:1988} from
comparison with earlier string
amplitude calculations (see e.g. \cite{Schwarz:1982}). In this paper we determine
the complete expression for the $\al'^2$ corrections to the
Born-Infeld part of a D9-brane
coupled to constant background fields at disk level in the string
coupling-constant perturbation expansion. We also determine $2n$-form
$2n$-derivative ($\al'^{n}$) corrections to the Wess-Zumino
part. By T-duality (dimensional
reduction), all
lower-dimensional cases can be obtained although we do not study this
issue in detail.  

In the next section we review some results in the literature for
future reference and also 
introduce our notation. In section \ref{sectWZ} we calculate 
$2n$-form $2n$-derivative corrections to the Wess-Zumino term and in section \ref{sectBI} we
calculate the $\al'^2$ corrections to the Born-Infeld action and also
show that the four-$F$ part correctly reproduces the previously known result. In section
\ref{sectOm} we then move on to
relate our results to what is known in the literature about the
lower-dimensional cases. In particular, we show that by dimensional
reduction of the Wess-Zumino term to $p{+}1$ dimensions, the part involving
only $\pa X$ (i.e. setting $F$ in the lower dimension to zero)
correctly reproduces the corrections to the Wess-Zumino term involving
the second fundamental form. In the final section we
make the observation that the four-index tensor in terms of which the
corrections determined in sections \ref{sectWZ} and \ref{sectBI} can
be 
expressed is reminiscent of the usual expression for the Riemann tensor. We make this idea
precise by showing that the tensor can be identified with the Riemann tensor
for the non-symmetric metric $h = \de + F$, where $\de$ is the flat
metric. In the appendices we list our conventions and give some
technical details of the calculations.

\setcounter{equation}{0}
\section{Preliminaries} \label{sectPre}
In this section we will review some known facts to introduce our
approach and some terminology. We will focus on the case of the D9-brane in
the type \II B theory. Comments on the lower-dimensional cases will
appear in a later section. 
The boundary conditions for an open string ending on a D9-brane are:  
$P_{\mu}=0$ and, for the fermionic
coordinates, $\bar{\Psi}^{\mu}=0$ (see
appendix \ref{convapp} for a description of our conventions). The
boundary state $\B$ is a BRST-invariant state
which imposes these 
boundary conditions, i.e. it satisfies $P_{\mu}\B =0$ and
$\bar{\Psi}^{\mu}\B=0$.
In the presence of a background gauge field, $A_{\mu}(X)$, with field
strength $F_{\mu\nu}(X) = \frac{\pa}{\pa X^{\mu}} A_{\nu}(X) -
\frac{\pa}{\pa X^{\nu}}A_{\mu}(X)$, the boundary condition requirements on
the boundary state $|B(F(X))\rangle$ are replaced by
\cite{Callan:1988}\footnote{Here we have absorbed a factor of
$2\pi\al'$ into $A_{\mu}$; what we call $F$ is thus really $2\pi\al'F$.}
\bea \label{bcB}
(2\pi\al'P_{\mu} + F_{\mu\nu}(X)\pa_{\si}X^{\nu})|B(F(X))\rangle = 0 \,,\non \\
(\bar{\Psi}_{\mu}- F_{\mu\nu}(X)\Psi^{\nu})|B(F(X))\rangle = 0 \,.
\eea
When restricting to the zero-mode part of $F(X)$, i.e.~$F(x)$, these
conditions can be written in terms of oscillators as
\bea \label{bcBcomp}
{}[(1{-}F)_{\mu\nu} \al_{n}^{\nu} +
(1{+}F)_{\mu\nu}\tilde{\al}_{-n}^{\nu}]\BF = 0 \,, \non \\
{}[(1{-}F)_{\mu\nu}d_{n}^{\nu} - i \eta
(1{+}F)_{\mu\nu}\tilde{d}_{-n}^{\nu}]\BF_{\mathrm{R}} = 0 \,,\non \\
{}[(1{-}F)_{\mu\nu}b_{r}^{\nu} - i \eta
(1{+}F)_{\mu\nu}\tilde{b}_{-r}^{\nu}]\BF_{\mathrm{NS}} = 0 \,.
\eea
Here we have introduced the notation $\BF:=|B(F(x)\rangle$. In
(\ref{bcBcomp}) $\eta\in\{+,-\}$ is the spin structure and the
subscripts R and NS refer to the Ramond and Neveu-Schwarz sectors, respectively.

A formal BRST-invariant solution to the equations (\ref{bcB}) is
\cite{Hashimoto:1999a,Hashimoto:1999b}
\be \label{BFX}
|B(F(X))\rangle=e^{-\frac{i}{2\pi \al'}\int
d\si \left[ \pa_{\si}X^{\mu}A_{\mu}(X) -
\frac{1}{2}\Psi^{\mu}\Psi^{\nu}F_{\mu\nu}(X)\right] }\B.
\ee 
This expression can be
succinctly written in terms of the superfield $\phi^{\mu} = X^{\mu} +
\tha\Psi^{\mu}$ and the super-covariant derivative $D = 
\tha\pa_{\si} -\pa_{\tha}$, as $|B(F(X))\rangle=e^{-\frac{i}{2\pi \al'}\int
d\si \D\tha D \phi^{\mu}A_{\mu}(\phi) }\B$.  As will become clear
later, we will not need the explicit form of $\B$ in this paper; the
properties (\ref{bcB}), (\ref{bcBcomp}) are sufficient for our purposes. 

It is known \cite{Callan:1988} that (in the absence of the NSNS
$B$-field) the parity-conserving (``Born-Infeld'') contribution (including derivative
corrections) 
to the effective action is  proportional to 
$\langle 0 |B(F(X))\rangle_{\mathrm{NS}}$, i.e.
\be \label{BIBFX}
 S_{\mathrm{BI}} \propto \langle 0 |e^{-\frac{i}{2\pi\al'} \int \D \si \D \theta D \phi^{\mu} A_{\mu}(\phi) } \B_{\mathrm{NS}} \,.
\ee
This expression is the operator analogue of the path integral
expression in the partition function approach:
\be 
S_{\mathrm{BI}} \propto \int \D
[X] \D[\Psi] e^{-\frac{i}{2\pi\al'} \int \D \si \D \theta D \phi^{\mu}
A_{\mu}(\phi)}\,.
\ee 
As indicated by the notation, the part of the effective action obtained from (\ref{BIBFX}) receives contributions only from the
NS sector as we will see below. The coupling to a constant $B$-field can be
obtained by observing that the only effect of the addition of a constant
$B$-field to the string $\si$-model action is to shift
$F$ to $F+B$. Another method to determine the (linear) couplings to the
background fields is to calculate the overlap of the boundary state
with the state corresponding to the relevant background field. For the case of a constant $B$-field the
result of the two approaches will be the same. The latter method is, however, more complicated
for the cases treated in this paper. The dependence on a constant
dilaton, $\Phi$, can also easily be determined: the only effect (in the string
frame) will be an overall 
multiplicative factor $e^{-\Phi}$, which can be understood from the
fact that we discuss disk-level contributions in the string
coupling perturbation expansion.

It is known that the
parity-violating part of the effective action involving the linear
couplings 
to the background RR form fields is obtained by
calculating the overlap between the state, $|C\rangle$, representing these form
fields, and the boundary state in the R sector (see e.g. \cite{DiVecchia:1999c}),
i.e. 
\be \label{WZBFX}
S_{\mathrm{WZ}} \propto \langle C|B(F(X)\rangle_{\mathrm{R}}\,.
\ee 

When neglecting derivatives acting on $F_{\mu\nu}$, we can replace $\int
\D\si\D\tha D \phi^{\mu} A_{\mu}(\phi)$ with  
$-\frac{1}{2} \int \D\si\D\tha D \phi^{\mu} \phi^{\nu}F_{\mu\nu}$. The lowest order contribution
to the parity-conserving part of the effective action is thus 
\be \label{BIBF}
S_{\mathrm{BI}}^{(0)} \propto \langle 0|B(F)\rangle_{\mathrm{NS}}\,,\; \mathrm{where}\;\; |B(F)\rangle_{\mathrm{NS}}:= e^{\frac{i}{4\pi\al'} \int \D \si \D \theta D \phi^{\mu} \phi^{\nu}F_{\mu\nu}}  |B\rangle_{\mathrm{NS}}\,.
\ee
A simple way to determine $S_{\mathrm{BI}}^{(0)}$ is to differentiate (\ref{BIBF}) with respect to
$F_{\mu \nu}$. This will bring down a factor $\frac{i}{4\pi\al'}
\int \D \si \D \theta D \phi^{\mu} \phi^{\nu}$ from the exponent. In
this expression, only the non-zero-mode part of the $\phi$'s,
$\tphi^{\eta}:=\phi^{\eta}-x^{\eta}$, contributes. The
next step is to use the
properties of $|B(F)\rangle$ (\ref{bcBcomp}) to turn all oscillators in $\tphi^{\nu}$ into creation
operators; we will denote the resulting expression
$\tphi^{\nu}_{+}$. Finally, one 
 moves $\tphi^{\nu}_{+}$ past $D \tphi^{\mu}$ and annihilates it
against $\langle 0|$ and in the process pick up a commutator $[D
\tphi^{\mu}(\si_1,\tha_1),\tphi_{+}^{\nu}(\si_{2},\tha_{2})]_{1\rar
2}$. We introduce the notation
\be \label{Gdef}
G_{12}^{\la\mu} := [ \tphi^{\la}(\si_1,\tha_1),\tphi_{+}^{\mu}(\si_{2},\tha_{2})] = D^{\la\mu}_{12} - \tha_1\tha_2 K_{12}^{\la\mu}\,.
\ee
In components, $D_{12}^{\mu\nu} = [ \tX^{\la}(\si_1),\tX_{+}^{\mu}(\si_{2})]$ and $K_{12}^{\mu\nu} = \{\tilde{\Psi}^{\mu},\tilde{\Psi}^{\nu}_{+}\}$. 
An explicit calculation leads to divergent expressions which need to be regularised. In this paper we will use the following regularisation 
\bea \label{DKprop}
D_{12}^{\mu\nu} &=& 2 \al' \sum_{n=1}^{\infty} \Big[ \frac{e^{-\ep n}}{n} (h_{S}^{\mu\nu} \cos[n(\si_2 - \si_1)] + i h_{A}^{\mu\nu}\sin[n(\si_2 - \si_1)] )\Big] \non \\
&=&  \al' \sum_{n=1}^{\infty} \Big[ \frac{e^{-\ep n}}{n} (h^{\mu\nu} e^{i n(\si_2-\si_1)} + h^{\nu\mu}e^{-in(\si_2-\si_1)})\Big]\,, \non \\
K_{12}^{\mu\nu}  &=& 2 \al' \sum_{r>0} \Big[ e^{-\ep r} (-h_{S}^{\mu\nu} \sin[r(\si_2 - \si_1)] + i h_{A}^{\mu\nu}\cos[r(\si_2 - \si_1)] )\Big] \non \\
&=&  i \al' \sum_{r>0}^{\infty} \Big[ e^{-\ep r} (h^{\mu\nu} e^{i r(\si_2-\si_1)} - h^{\nu\mu}e^{-ir(\si_2-\si_1)})\Big] \,,
\eea
where $\ep{>}0$ is the regulator. In (\ref{DKprop}) we have used the notation $h^{\mu\nu} = \left(\frac{1}{1 +
F}\right)^{\mu\nu}$ for the inverse of $h_{\mu\nu}:=
(1{+}F)_{\rho\nu}$, (i.e., $h^{\mu\nu}h_{\nu\rho}=\de^{\mu}_{\rho}$ and
$h_{\rho\nu}h^{\nu\mu}=\de^{\mu}_{\rho}$), and also $h_{A}^{\mu\nu} =
\frac{1}{2}[h^{\mu\nu} - h^{\nu\mu}]$ and $h_{S}^{\mu\nu} =
\frac{1}{2}(h^{\mu\nu} + h^{\nu\mu})$.
In the expression for $K_{12}^{\mu\nu}$, $r$ runs over the (positive)
half integers in the NS sector, whereas it takes (positive) integer values in the R sector. 
The results (\ref{DKprop}) agree with the usual expressions for the $F$-dependent propagators 
on the boundary of the disk \cite{Abouelsaood:1987,Andreev:1988}. The spin structure parameter $\eta$ drops out of the propagator, so the GSO projection (sum over spin structures) can be absorbed into the zeroth order boundary state and will not enter in our discussion. 

We now return to the determination of $S_{\mathrm{BI}}^{(0)}$. From the above results it
follows that 
$ \int \D \tha_1 [D\phi^{\mu}(\si_1,\tha_1),\phi_{+}^{\nu}(\si_{2},\tha_{2})]_{1\rar
2} = -2i \al' h_{A}^{\mu\nu}[\sum_n e^{-\ep n} - \sum_r e^{-\ep r}]$,
which when the regulator is removed becomes $i \al' h_{A}^{\mu\nu}$. 
 From the integration over $\si$ we get an additional factor of $2\pi$.  Thus, we obtain the following differential equation: $\frac{\pa
S_{\mathrm{BI}}^{(0)} }{\pa F_{\mu\nu}} =- \frac{1}{2}h_{A}^{\mu\nu}S_{\mathrm{BI}}^{(0)}$, for the lowest order term, which (up to a multiplicative
constant, which is identified with minus the tension, $T_9$) has the
usual Born-Infeld action
\be \label{SBI0}
S^{(0)}_{\mathrm{BI}} = -T_9 \int \D^{10}x \sqrt{\det(1+F)} = -T_9\int \D^{10} x \sqrt{\det\, h }\,,
\ee
as its unique solution. This is most easily proven by going to the special
Lorentz frame where $F_{\mu\nu}$ is block diagonal. 

The lowest order contribution to the Wess-Zumino term was calculated in
detail in \cite{DiVecchia:1999c} using the present framework. Only the
zero-mode part of $\BF$ contributes (since
$[D_1G_{12}^{\mu\nu}]_{1\rar2}=0$ in the R sector) and the result is the well known expression 
\be \label{SWZ0}
S^{(0)}_{\mathrm{WZ}} = T_p \int C\we e^F\,,
\ee 
where $C$ is the sum of the RR potential forms. 

To study derivative corrections to the effective action we will
expand $A_{\mu}(\phi)$ around the zero mode of $\phi$, i.e. we split $\phi$ as $\vphi_0 +
\tilde{\phi}$, and Taylor expand: 
\be \label{Aexp}
A_{\mu}(\phi) = \sum_{n=0}^{\infty}
\frac{1}{n!}\tilde{\phi}^{\nu_1}\cdots\tilde{\phi}^{\nu_n} \pa_{\nu_1} \cdots \pa_{\nu_n} A_{\mu}(\vphi_0) \,.
\ee
 This procedure leads to a well defined
perturbative expansion in terms of the number of derivatives.
Inserting the expansion (\ref{Aexp}) into $\int \D \si \D \tha
D_{\si}\phi^{\mu}A_{\mu}(\phi)$ results in
\bea \label{Fexp}
&& -\,\int \D \si \D\tha 
\sum_{k=0}^{\infty} \ts{\frac{1}{(k{+}1)!}\frac{k{+}1}{k{+}2}}D \tphi^{\nu}\tphi^{\mu}\tphi^{\la_1} \cdots
\tphi^{\la_k}\pa_{\la_1} \cdots \pa_{\la_k} F_{\mu\nu}(x) \non \\ &&- \, \int \D
\si [\tilde{\Psi}^{\mu}\psi_0^{\nu} + \half\psi_0^{\mu}\psi_0^{\nu}] \sum_{k=0}^{\infty} \frac{1}{k!}\tX^{\la_1} \cdots
\tX^{\la_k}\pa_{\la_1} \cdots \pa_{\la_k} F_{\mu\nu}(x)\,.
\eea
In the NS sector $\vphi^{\mu}_0 = x^{\mu}$, whereas in the R sector
 $\vphi^{\mu}_0 = x^{\mu} + \tha \psi_0$. Hence, in the NS sector the second term in (\ref{Fexp})
 is absent. The $k=0$ terms in the above expansion give rise to the
 $F(x)$ dependence in  $\BF$. 
 As discussed in \cite{Hashimoto:1999a,Hashimoto:1999b} the boundary
state (\ref{BFX}) is formal in the sense that the perturbative expansion it
generates contains divergences
which have to be regularised. There are two ways to view these
divergencies. Either one takes the view that the $F$ one started with, i.e. the
one appearing in the boundary state, is the ``true'' $F$. 
The terms multiplying the divergent terms then have to vanish. These
 conditions are closely
related to the vanishing of the
$\beta$-function in the $\beta$-function approach as discussed in \cite{Hashimoto:1999a,Hashimoto:1999b}. The
other way  to view the divergencies is to say that one started with the wrong
$F$, and to take this fact into account by
making field redefinitions to absorb the divergencies (since the theory is
renormalisable this can always be done). We will
take the latter viewpoint in this paper. The effective D-brane action
 is thus obtained from (\ref{BIBFX}) and (\ref{WZBFX}) after renormalisation.  
At least for the parity-conserving part one can also reformulate the algebraic operator approach used so far in terms of Feynman diagrams. Since
$F$ satisfies the Bianchi identity, $\pa_{[\la}F_{\mu\nu]}=0$, there will be relations between
different contributions which reduce the rather large number of diagrams
and simplifies the analysis. We found it easier to take such relations
into account in the operator language. 

We will now briefly review, using the boundary state
formalism, what is known about the $\cO(\al')$ corrections to the
effective action for a D-brane or, stated in another way, the corrections
involving two additional derivatives compared to the zeroth order
terms (\ref{SBI0}), (\ref{SWZ0}).   
To this end we only need to keep the first three terms ($k=0,1,2$) in
the expansion (\ref{Fexp}), insert them into the exponential in (\ref{BFX})
and expand (again retaining only the terms with at most two
derivatives). We only need to consider terms with an even number of
derivatives, since it is not possible to form a scalar out of
$F_{\mu\nu}$ and an odd number of derivatives.
 
Let us start by discussing the Wess-Zumino part. 
 The two-derivative corrections to the Wess-Zumino term are given by $\langle
 C|B(F(X)\rangle_{\mathrm{R}}$, with $|B(F(X)\rangle_{\mathrm{R}}$
expanded to second order in derivatives. In the various correlation
functions that arise we let, as before,  
 the $\tphi^{\mu}$ to the far right act on the boundary state $\BF$ to turn
 annihilation operators into creation operators using the property
 (\ref{bcB}) and then move it to the left and finally annihilate it
 against $\langle C|$. In the process we will pick up commutators of
the form (\ref{Gdef}). Continuing this process gives an expression for
the contribution to the action expressed in terms of the propagator, $G$.
The corrections can be classified according to how many
zero modes the expressions from which they arise contains (which, as will be clear later, in
turn translates into forms of different degrees). We will now show that the part involving no zero modes
vanishes to any order in the derivative perturbation theory. More
precisely, we will show that 
\be 
\langle C |(\int \D \tha_1 D_1 \tphi
\, \tphi^{k_1})(\int \D\tha_2 D_2\tphi \,\tphi^{k_2}) \cdots (\int \D\tha_n
D_n\tphi \,\tphi^{k_n}) \B =0\,,
\ee
where we have suppressed all indices.  We start by considering $D_1\tphi$. We get zero if this term is
 contracted with a $\tphi$ within the same factor since $[D_1
G_{12}]_{2\rar1}=0$ (which follows from the fact that in the R sector $K^{\la\mu}_{12} =
 \pa_{2}D^{\la\mu}_{12} = -\pa_1 D^{\la\mu}_{12}$). Assume therefore that $D_1 \tphi$ is contracted with a $\tphi$ in the
$m$th factor (by integration by parts we can always arrange it to be one of
the  $\tphi$'s rather than $D_m\tphi$). Now consider $D_m\tphi$. As
before we can not contract it with a $\tphi$ within the same
factor. Also, if it is contracted with a $\tphi$ in the first factor we get zero since
$D_1 G_{1m}D_m G_{1m} \propto (\tha_1-\tha_m)^2 = 0$. Thus we can
assume that $D_m \tphi$ is contracted with a $\tphi$ in the $n$th factor. Now look
at $D_n \tphi$. There are two possibilities. 1) it is contracted with
a $\tphi$ in the first factor. In this case we select a  $D_k
\phi$ from another factor and start over. 2) it is contracted with the
$r$th term. In this case we continue as before. It should be clear
that the procedure outlined above leads to a result proportional to
$\prod_{p,q} D_{p}G_{pq} \propto \prod_{p,q} (\tha_p -\tha_q)$, for
certain $p,q$. Since
there are $j$ factors this has in turn to be proportional to $\prod_{i=1}^{j}
\tha_i$, but since $\prod_{p,q} (\tha_p -\tha_q)$ is translational
invariant whereas $\prod_{i=1}^{j}
\tha_i$ is not, the proportionality constant has to be zero. We thus conclude that there
are no derivative corrections to the ``zero-form'' part of the Wess-Zumino term. By the same argument
there can arise no scalar functions (constructed out of $h^{\mu\nu}$'s
and $\pa^k F$'s) multiplying the corrections with
form degrees $>0$. 
 
Let us now return to the discussion of the $\cO(\al')$ corrections to the
Wess-Zumino term. The two
contributions to the $(\psi_0)^2$ part can be removed by the field
redefinition  \cite{Hashimoto:1999b}
\be \label{deF1}
\de F_{\mu\nu} = \al' (\ln \ep \, h^{\la\rho}\pa_{\la}\pa_{\rho}F_{\mu\nu} - 2 \ln (2\ep) h^{(\rho|\eta} \pa_{[\mu} F_{|\eta\de}h^{\de|\la)}\pa_{\la}F_{\rho|\nu]} )\,.
\ee 
The $\ln \ep$ terms can further be absorbed into a field redefinition of
$A_{\mu}$ as 
\be \label{deA1}
\de A_{\mu} = \al' \ln \ep \,h_{S}^{\la\rho}\pa_{\la}F_{\rho\nu} \,,
\ee
but the remaining $\ln 2$ term cannot be absorbed into a redefinition
of $A_{\mu}$. The easiest way to see this is to note that the
corresponding $\de F_{\mu\nu}$ is not closed, i.e. $\D \de F_{\mu\nu}
\neq 0$. The $\ln
2$ term is however renormalisation scheme dependent and there exists schemes in
which it vanishes. One example of such a scheme is the one discussed in
\cite{Andreev:1988}. In this scheme the propagator is replaced by
\bea
D_{12}^{\mu\nu} &=& 2 \al' \sum_{n=1}^{\infty} \Big[ \frac{e^{-\ep n}}{n} (h_{S}^{\mu\nu}(F_n) \cos[n(\si_2 - \si_1)] + i h_{A}^{\mu\nu}(F_n)\sin[n(\si_2 - \si_1)] )\Big] \non \\
&=&  2\al' \sum_{n=1}^{\infty} \Big[ \frac{e^{-\ep n}}{n} (h^{\mu\nu}(F_n) e^{i n(\si_2-\si_1)} + h^{\nu\mu}(F_n)e^{-in(\si_2-\si_1)})\Big] \,,
\eea 
where $(F_n)_{\mu\nu} := e^{-\ep n}F_{\mu\nu}$. The expression for
$K^{\mu\nu}$ is similar.  
 Another scheme is $\ze$-function regularisation
in which both $\ln\ep$ and $\ln(2\ep)$ get replaced by $\ze(1)$ \cite{Hashimoto:1999b}. The
$\ze$-function regularisation method 
becomes less useful at higher loops as the sums become more
complicated; we will therefore not use it in this paper.  

The $\cO(\al')$ corrections to the Born-Infeld part of the action is obtained
from (\ref{BIBFX}) with (\ref{BFX}) expanded to second order in
derivatives. The calculations can be found in \cite{Andreev:1988} (see
also \cite{Hashimoto:1999b}). The resulting expression can be completely
removed by the same field redefinition, (\ref{deF1}), as for the
Wess-Zumino term.

\setcounter{equation}{0}
\section{Corrections to the Wess-Zumino term} \label{sectWZ}

In this section we will study derivative corrections involving the
world-volume gauge field to the Wess-Zumino term. More precisely, we will determine
 corrections which couple linearly to the Ramond-Ramond fields.

Let us first note that we need an even number of fermionic zero modes (or
equivalently an even number of $\tilde{\Psi}'s$) to get a
non-vanishing result. As is explained in appendix \ref{techapp}, this translates into the statement that only corrections of even form degrees are present.  Furthermore, the contribution from the zero-mode
independent part vanishes (to all orders in the derivative expansion) as a consequence of the general result given
in the previous section, thus there are no zero-form corrections. 

Let us start by looking at the $\al'^2$ corrections. We will first discuss the $(\psi_0)^2$ terms. These
terms can potentially be removed by a
field redefinition $\de F_{\mu\nu}$ of order $\al'^2$, since the
zeroth order term transforms as $\de C e^F = C e^F \de F$, which, as
explained in appendix \ref{techapp}, is of
the same form as the corrections arising from the $(\psi_0)^2$
terms. Indeed, this is precisely what happens (see appendix
\ref{techapp} for details). An independent argument showing that it
has to happen is the following: the zeroth order Wess-Zumino term
satisfies the 
Bianchi identity $\D [C e^{F}] = R e^{F}$, where (when
$B=\mathrm{constant}$) $R = \D C$. In order for the Wess-Zumino term
to still satisfy a Bianchi identity in  the presence of two-form
corrections, $C e^{F} W_2$, the two-form $W_2$ has to be closed. This
means that the two-form correction can be removed by choosing $\de F =
- W_2$, which, since $W_2$ is closed,  translates (at least locally)
into $\de A = -w_1$, where $\D w_1 = W_2$, thus showing that $W_2$ can
be removed by a field redefinition of $A_{\mu}$. This argument works
at any order, showing that the two-form corrections can be removed by a field redefinition. This singles out a particular action, and is a natural way to ``gauge'' fix the field redefinition ambiguity.  

The divergent $(\psi_0)^4$ terms at order $\al'^2$ must be cancelled by
the terms arising from inserting $\de F_{\mu\nu}^{(1)}$ given in
(\ref{deF1}) into the zeroth order action and expanding to second order.
That this is works out correctly is a
non-trivial check. The remaining finite terms give rise to
corrections that can be expressed in terms of the tensor 
\be 
S_{\rho_1\rho_2 \mu_1\mu_2} := 
\pa_{\rho_1}\pa_{\rho_2}F_{\mu_1\mu_2} + 2h^{\nu_1 \nu_2}\pa_{\rho_1}
F_{[\mu_1|\nu_1} \pa_{\rho_2|}F_{\mu_2]\nu_2}\,.
\ee
Since $S_{\rho_1\rho_2\mu_1\mu_2}$ is antisymmetric in its last two
indices it can be considered as a differential two-form with two
additional indices which we will denote $\mathsf{S}_{\rho_1\rho_2} :=
\ts{\frac{1}{2!}}S_{\rho_1\rho_2\mu_1\mu_2}\D x^{\mu_1}\we \D
x^{\mu_2}$. With this definition, the action including the $\al'^2$
corrections can be written (see appendix \ref{techapp} for details)
\be \label{WZact}
S^{(2)}_{\mathrm{WZ}} = T_{9} \int C\we e^{F} \we(1 + \frac{(2\pi \al')^2}{48}h^{\rho_4\rho_1}h^{\rho_2\rho_3} \mathsf{S}_{\rho_1\rho_2}\we \mathsf{S}_{\rho_3\rho_4})\,.
\ee
One can show that the correction satisfies $\D (h^{\rho_4\rho_1}h^{\rho_2\rho_3}
\mathsf{S}_{\rho_1\rho_2}\we \mathsf{S}_{\rho_3\rho_4}) = 0$, i.e. it
is closed, which as discussed above is important for consistency.

It is possible to generalise the 4-form 4-derivative corrections
determined above and calculate the corresponding $2n$-form $2n$-derivative corrections to the Wess-Zumino term. Details can be found in appendix \ref{techapp} and the result is 
\be \label{2nform}
S = T_9 \int C\we e^{F} \we(1 + \ds{\sum_{k=2}^{5}} W_{2k})\,,
\ee
where (wedge products are suppressed) 
\bea \label{Ws}
W_4 &=& (\al')^2 \ts{\frac{\ze(2)}{2}} h^{\rho_4\rho_1}h^{\rho_2\rho_3} \mathsf{S}_{\rho_1\rho_2} \mathsf{S}_{\rho_3\rho_4}\,, \non \\
W_6 &=& (\al')^3 \ts{\frac{\ze(3)}{3}}
h^{\rho_6\rho_3}h^{\rho_4\rho_1}h^{\rho_2\rho_5}
\mathsf{S}_{\rho_1\rho_2} \mathsf{S}_{\rho_3\rho_4}
\mathsf{S}_{\rho_5\rho_6} \,, \non \\
W_8 &=& (\al')^4 [ \ts{ \frac{\ze(4)}{4} } 
h^{\rho_8\rho_5}h^{\rho_6\rho_3}h^{\rho_4\rho_1}h^{\rho_2\rho_7} 
\mathsf{S}_{\rho_1\rho_2} \mathsf{S}_{\rho_3\rho_4} 
\mathsf{S}_{\rho_5\rho_6} \mathsf{S}_{\rho_7\rho_8} \non \\ &&
+ \,\ts{\frac{1}{2}} \ts{ \left(\frac{\ze(2)}{2}\right)^2 }
(h^{\rho_4\rho_1}h^{\rho_2\rho_3} \mathsf{S}_{\rho_1\rho_2}
\mathsf{S}_{\rho_3\rho_4})(h^{\rho_8\rho_5}h^{\rho_6\rho_7}
\mathsf{S}_{\rho_5\rho_6} \mathsf{S}_{\rho_7\rho_8}) ] \,, \non \\
W_{10 }&=& (\al')^5 [ \ts{ \frac{\ze(5)}{5} } h^{\rho_{10}\rho_7} h^{\rho_8\rho_5}h^{\rho_6\rho_3}h^{\rho_4\rho_1}h^{\rho_2\rho_9} 
\mathsf{S}_{\rho_1\rho_2} \mathsf{S}_{\rho_3\rho_4} 
\mathsf{S}_{\rho_5\rho_6} \mathsf{S}_{\rho_7\rho_8}\mathsf{S}_{\rho_9\rho_{10}} \non \\ &&+ \,\ts{\frac{\ze(2)}{2}\frac{\ze(3)}{3}}(h^{\rho_4\rho_1}h^{\rho_2\rho_3} \mathsf{S}_{\rho_1\rho_2} \mathsf{S}_{\rho_3\rho_4})(h^{\rho_{10}\rho_7}h^{\rho_8\rho_5}h^{\rho_4\rho_9} \mathsf{S}_{\rho_5\rho_6}
 \mathsf{S}_{\rho_7\rho_8} \mathsf{S}_{\rho_9\rho_{10}})] \,,
\eea
and $\ze(n)$ is the Riemann $\ze$-function. Notice that
$4n{+}2$-form corrections are possible, which is not the case for the
corrections involving the bulk Riemann tensor and the second
fundamental form \cite{Green:1997,Bachas:1999}. This is
not a contradiction, since as we will see in a later section the
$4n{+}2$-form corrections vanish after dimensional reduction and
setting the gauge field in the lower dimension to zero, thus no
$4n{+}2$-form corrections involving only the transverse scalars appear. 

The corrections (\ref{2nform}), (\ref{Ws}) can be written even more
compactly if we  define the trace operation $\hat{\tr}$ over
$\mathsf{S}_{\rho_1\rho_2} \cdots \mathsf{S}_{ \rho_{2n{-}1}\rho_{2n}
}$ as in (\ref{Ws}). In words: start with the index to far right and
contract this index with the first index on the $\mathsf{S}$ next to
it to the left using $h^{ \rho_{2n}\rho_{2n-3} }$; then contract the
remaining index on $\mathsf{S}_{ \rho_{2n-3}\rho_{2n-2} }$ with the
first index on the next $\mathsf{S}$ and so on until $\mathsf{S}_{
\rho_1\rho_2 }$ is reached; finally contract its remaining index
($\rho_2$) with the remaining index ($\rho_{2n-1}$) on $\mathsf{S}_{
\rho_{2n-1}\rho_{2n} }$ using $h^{\rho_2\rho_{2n-1} }$. This
definition is cyclic, which justifies calling it a trace. Using
$\hat{\tr}$  the $2n$-form $2n$-derivative corrections (\ref{2nform}),
(\ref{Ws}) can be written 
\be 
S = T_9 \int C \exp [ F + \sum_{n=2}^{5} \ts{\frac{\ze(n)}{n}} (\al' \hat{\tr}\,\mathsf{S})^{n} ]\,.
\ee
The form of this expression is closely related to the $\Gamma$-function, as can be seen by recalling that 
\be
-e^{-\ga z} z \Ga(-z) = \exp({\ds{\sum_{n=2}^{\infty}} \ts{\frac{\ze(n)}{n}} z^{n} }) = \ds{\prod_{l=1}^{\infty}}(1{-}{\frac{z}{l}})^{-1}e^{-\frac{z}{l}}\,,
\ee
where $\ga$ is the Euler constant. 

Finally, let us remark that since the only effect of a constant NSNS
$B$-field in the string $\si$-model is to shift $F \rar F+B$, we can
easily modify our expressions to include such a field. To treat the
case with a non-constant $B$-field, the fact that string theory is
symmetric under the transformation  $B \rar B + \pa_\mu \La_{\nu} -
\pa_{\nu}\La_{\mu}$ should be taken into account. Under this transformation the gauge field on the
brane transforms as $A_{\mu} \rar A_{\mu} -\La_\mu$. The most
straightforward way to write invariant expressions is thus to replace
$F$ by 
$F+B$. However, this is not the end of the story, since in
addition $H=\D B$ is also invariant under the above symmetry and may appear explicitly.  To
treat the case with non-constant $B$-fields the results obtained in
\cite{Haggi-Mani:2000} may be useful.  Presumably,
there are also corrections  which are non-linear in the RR form fields,
$C_{p}$. For non-constant backgrounds there are also corrections
involving the bulk Riemann tensor
\cite{Green:1997,Bachas:1999}. 
In addition, corrections involving derivatives of
the dilaton $\Phi$, $B_{\mu\nu}$ and the $C_{p}$'s are
expected.

\setcounter{equation}{0}
\section{Corrections to the Born-Infeld term} \label{sectBI}
In this section we derive the complete expression (at disk level) for the bosonic
four-derivative  corrections to the
Born-Infeld part of the effective action for a D-brane coupled to constant background
fields. In particular, the expression involves all orders of the gauge field $F$. The
result is obtained via a 3-loop  calculation of the string $\si$-model
 partition function, using the boundary state operator language. The calculations are more involved
than the ones performed for the Wess-Zumino term in the previous section. The fermionic modes are now half-integer moded
and there are no fermionic zero modes. A complication is the presence
of renormalisation scheme dependent terms. Fortunately, we can use
the fact that in the course of the determination of the contribution
to the Wess-Zumino term one field redefinition was singled out, namely the one
that removed the $(\psi_0)^2$ (two-form) contribution. Precisely this
field redefinition will greatly simplify the corrections to the Born-Infeld part
as well. After a tedious calculation, described in more detail in appendix \ref{techapp}, one arrives at the following result 
\bea \label{BI2}
S_{\mathrm{BI}}^{(2)} &=& - T_{9} \int \D\si \sqrt{\det\, h }\Big[ 1 + 
\frac{(2\pi\al')^2}{96}\bigg( -h^{\mu_4\mu_1}h^{\mu_2\mu_3}h^{\rho_4\rho_1}h^{\rho_2\rho_3}S_{\rho_1\rho_2\mu_1\mu_2}S_{\rho_3\rho_4\mu_3\mu_4}
 \non \\ && \qquad \qquad \qquad\qquad \qquad \qquad \quad \;\;\;+\, \half
h^{\rho_4\rho_1}h^{\rho_2\rho_3}S_{\rho_1\rho_2}S_{\rho_3\rho_4}\bigg)
\Big]\,,
\eea
where as before
\be 
S_{\rho_1\rho_2 \mu_1\mu_2} =
\pa_{\rho_1}\pa_{\rho_2}F_{\mu_1\mu_2} + 2h^{\nu_1 \nu_2}\pa_{\rho_1}
F_{[\mu_1|\nu_1} \pa_{\rho_2|}F_{\mu_2]\nu_2}\,,
\ee
and in addition 
 $S_{\rho_1\rho_2}:=h^{\mu_1\mu_2}S_{\rho_1\rho_2\mu_1\mu_2}$ (not to
 be confused with the two-form $\mathsf{S}_{\rho_1\rho_2}$ which
 appeared in the previous section). It is straightforward to check
 that the result (\ref{BI2}) only contains even powers of $F$, by using
 $h^{\mu\nu}(-F)= h^{\nu\mu}(F)$ together with
 $S_{\rho_1\rho_2\mu_1\mu_2}(-F) = S_{\rho_2\rho_1\mu_2\mu_1}(F)$.

The four-derivative corrections to the Born-Infeld part of the action
involving four $F$'s and four derivatives have been known
for a long time. These corrections can be extracted from the four-string
scattering amplitude (see e.g. \cite{Schwarz:1982}) and have the form \cite{Andreev:1988}
\bea \label{t8F4}
S_{ (\pa F)^4 } &=& -T_9 \frac{\al'^2 \pi^2}{24} \int \Big[ \ts{
\frac{1}{8} }\tr (\pa_{\rho} F \pa^{\rho} F )\tr(\pa_{\la} F \pa^{\la}
F)  + \ts{\frac{1}{4} }\tr (\pa_{\rho} F \pa_{\la} F )\tr(\pa^{\rho} F
\pa^{\la} F ) \non \\ &&\qquad \qquad \qquad \;\, -\,
\ts{\frac{1}{2}}\tr(\pa_{\la} F \pa_{\rho} F \pa^{\la} F \pa^{\rho} F) - \tr(\pa_{\la} F \pa_{\rho} \pa^{\rho} F \pa^{\la} F) \Big] \,.
\eea
The integrand can be rewritten in terms of the well known eight-index $t_8$-tensor as $-\frac{1}{16}t_8 \pa_{\rho} F \pa_{\la} F \pa^{\la} F
\pa^{\rho} F$  (the definition of 
$t_8$ is given in appendix \ref{techapp}). To check whether our result
(\ref{BI2}) correctly reduces to (\ref{t8F4}) when restricting to the $\cO(F^4)$
terms is rather involved, since the two expressions need only be equal
up to terms which can be removed by field redefinitions. (This
ambiguity also accounts for the seeming discrepancy between
(\ref{t8F4}) and the expression in \cite{Andreev:1988}.) It turns out
that the four different types of terms entering in (\ref{t8F4}) form a basis for any term constructed out of four
derivatives and four $F$'s modulo field redefinitions, integrations by
parts and use of the Bianchi identity. This fact is shown in appendix
\ref{techapp}. Armed with this result it is fairly straightforward to
show that the two expressions do indeed  agree. This is a non-trivial
check of our result. For more details see appendix
\ref{techapp}. Another test of our result is to check that it
transforms correctly under the Seiberg-Witten map
\cite{Seiberg:1999}. We hope to return to this question elsewhere
\cite{Wyllard:2xxx}. 

As before, the dependence on constant background fields can be taken
care of by replacing $F$ by $F+B$ and (in the string frame) introduce
an overall factor of $e^{-\Phi}$. So far we have only discussed
derivative corrections at disk level in the string-coupling 
perturbation expansion. In addition there should also be contributions
from loops and non-perturbative effects. Such corrections were discussed
in \cite{Green:2000}
and were argued  to appear for the 
$(\pa F)^4$ part of the action for the D3-brane as an overall multiplicative
function. This is presumably also true for our four-derivative result involving all orders
of $F$.

\setcounter{equation}{0}
\section{Lower-dimensional cases} \label{sectOm}

The results in sections \ref{sectWZ} and \ref{sectBI} are valid for the nine-brane case. To obtain the
corrections to the lower-dimensional branes we need to dimensionally
reduce the above action (T-duality). We will now change notation slightly and label 
ten-dimensional indices by $M,N,\ldots$, world-volume indices by
$\mu,\nu,\ldots$, and indices in the directions normal to the brane by $i,j,\ldots$. The
embedding of the brane into the ten-dimensional target space is described by the elements $\pa_{\mu}Y^{M}$, which span a
local frame of the tangent bundle and the objects $\xi^i_M$ which play a similar
role for the normal bundle (for more details, see e.g. \cite{Bachas:1999}). World-volume indices are raised and lowered
with the induced metric $g_{\mu\nu} :=
\de_{MN}\pa_{\mu}Y^{N}\pa_{\nu}Y^{M}$, indices in the directions
normal to the brane with $\de_{ij}$ and ten-dimensional indices with
$\de_{MN}$. 

The fundamental quantity describing the embedding is the
second fundamental form, $\Om^{M}_{\mu\nu}$, defined as
the covariant derivative of $\pa_{\mu}Y^{M}$; for a flat background we
thus have
\be
\Om^{M}_{\mu\nu} = \pa_{\mu}\pa_{\nu}Y^{M} -
{\Ga_{\!\!\mathrm{T}}}^{\la}_{\mu\nu}\pa_{\la}Y^{M} \,,
\ee
where ${\Ga_{\!\!\mathrm{T}}}^{\la}_{\mu\nu}$ is the connection constructed
from the induced metric. More explicitly, we find $\Om^{M}_{\mu\nu} =
(\de^{M}_{L} - \pa_{\la} Y^{M}
g^{\la\rho}\pa_{\rho}Y^{K}\de_{KL})\pa_{\mu}\pa_{\nu}Y^{L} =:
P^{M}_{L}\pa_{\mu}\pa_{\nu}Y^{L}$. It can easily be shown that $P^{M}_{L}$ is a projection
matrix, $P^{M}_{N} P^{N}_{L} = P^{M}_{L}$.  Since $\de^{MN} =
\pa_{\la}Y^{M}g^{\la\rho}\pa_{\rho}Y^{N} +
\xi^{M}_{i}\de^{ij}\xi^{N}_{j}$, we find that
$\xi^{M}_{i}\de^{ij}\xi^{N}_{j} = P^{MN}$. Is is easy to check that the projection of
$\Om^{M}_{\mu\nu}$ onto the tangent
bundle, $\pa_{\rho}Y^{N}\de_{NM}\Om^{M}_{\mu\nu}$, vanishes. We
introduce the notation $\hat{\Om}^{i}_{\mu\nu} :=
\xi^{i}_{M}\Om^{M}_{\mu\nu}$ for the second fundamental form projected
onto the normal bundle.

The $\al'^2$ corrections  to the Wess-Zumino term involving the
Riemann tensor are \cite{Green:1997,Bachas:1999}
\be
S_{R^2} = T_{p} \int C\we e^{F} \we \frac{(2\pi \al')^2}{96}  [\tr (R_{\mathrm{T}} \we R_{\mathrm{T}}) - \tr(R_{\mathrm{N}} \we R_{\mathrm{N}}) ] \,,
\ee
where the subscripts T and N refer to the normal and tangent bundles,
respectively. In a flat background we have the relations
\bea \label{flatR}
(R_{\mathrm{T}})_{\mu\nu\rho\la} &=& \de_{ij}\hat{\Om}^i_{\mu\rho}\hat{\Om}^{j}_{\nu\la} - (\mu\leftrightarrow \nu) \,,\non \\
{(R_{\mathrm{N}})_{\mu\nu}}^{ij} &=& g^{\rho\la}\hat{\Om}^i_{\mu\rho}\hat{\Om}^{j}_{\nu\la} - (\mu\leftrightarrow \nu)\,.
\eea
The above corrections are thus expressed in terms of the second
fundamental form. In the static gauge,
$Y^{\mu} = x^{\mu}$ and $Y^{i} = X^{i}(x)$, we find
\be \label{R22}
S^{(2)}_{R} = T_{p} \frac{(2\pi
\al')^2}{48} \int C e^{F}  P^{i_1 i_2}P^{i_3
i_4}g^{\rho_1\rho_4}g^{\rho_2\rho_3}\pa_{\rho_1}\D
X_{i_1}\pa_{\rho_2}\D X_{i_2} \pa_{\rho_3}\D X_{i_3}\pa_{\rho_4}\D X_{i_4}\,.
\ee
Next we will investigate whether these corrections
are correctly reproduced by the dimensional reduction of our result, 
(\ref{WZact}). When reducing, the gauge field $F_{MN}$ splits into $F_{\mu\nu}$ and $F_{\mu i} = \pa_{\mu}X_{i}$ ($F_{ij}\equiv 0$). Thus,
\be
h_{MN} = \left(\ba{cc} (1{+}F)_{\mu\nu} & \pa_{\mu}X_j \\ -\pa_{\nu}X_{i} & \de_{ij} \ea \right)\,.
\ee
The inverse, $h^{MN}$, becomes
\be
h^{MN} = \left(\ba{cc} \tilde{h}^{\mu\nu} & -\tilde{h}^{\mu\la}\pa_{\la}X^j \\ \pa_{\la}X^{i}\tilde{h}^{\la\nu} & \de^{ij} - \pa_{\la}X^{i}\tilde{h}^{\la\rho}\pa_{\rho}X^j \ea \right)\,.
\ee
Here $\tilde{h}^{\mu\nu} = (\frac{1}{\eta + F + \pa X^i \pa
X_i})^{\mu\nu}$. When $F_{\mu\nu}=0$, $\tilde{h}^{\mu\nu}$ reduces to
the inverse of the induced metric, $g^{\mu\nu}$. If we concentrate on
the part of $\mathsf{S} \wedge \mathsf{S}$ in (\ref{WZact}) which has four indices in the
world-volume directions, we
see that only the $(\pa F)^4$ part contributes when
$F_{\mu\nu}=0$. Furthermore, observing that (when $F$ is zero) $h^{ij} = P^{ij}$ we find 
complete agreement with (\ref{R22}). 

When reducing the zeroth order Wess-Zumino term from ten to nine dimensions
the part of $e^F$ with one index transverse to the world volume gets
reinterpreted as part of the pullback of $C$ to nine dimensions by
T-duality \cite{Green:1996}. We have not investigated how this
statement generalises for the corrections considered in section
\ref{sectWZ}.

Let us now look at the higher-degree corrections. When $F_{\mu\nu}=0$,
one can easily show that  $S_{\rho_1\rho_2\mu_1\mu_2} = -S_{\rho_2\rho_1\mu_1\mu_2}$, which
implies that the six- and ten-form corrections in (\ref{Ws}) vanish 
when we put the gauge field to zero. The eight-form corrections are,
however, non-vanishing. These should be compared with the eight-form
corrections involving the second fundamental form. Starting from the
expression \cite{Green:1997}
\be
S_{R} = T_{p} \int C\we e^{F} \we \frac{\sqrt{ \hat{ \mathcal{A} }(4\pi^2
\al'R_{\mathrm{T}}) } }{\sqrt{\hat{\mathcal{A}} (4\pi^2 \al'R_{\mathrm{N}})}}\,,
\ee
where 
\be
\sqrt{\hat{\mathcal{A} }(4\pi^2 \al'R)}=1 -
\ts{\frac{\pi^2\al'^2}{24}}\tr{R^2} -
\ts{\frac{\pi^4\al'^4}{720}}\tr(R^4) + \ts{\frac{\pi^4\al'^4}{192}}(\tr(R^2))^2
+ \ldots\,,
\ee
and using the results (\ref{flatR}) it is straightforward to check
that also the eight-form corrections are correctly reproduced. 
In \cite{Bachas:1999} $\al'^2$ corrections to the Born-Infeld part of
the action involving the second
fundamental form  were discussed. It should also
be possible to check if these corrections are correctly
reproduced by our result.

\setcounter{equation}{0}
\section{Connection to non-symmetric gravity}

In the corrections obtained in sections \ref{sectWZ} and \ref{sectBI} only the combination $h_{\mu\nu} = (1{+}F)_{\mu\nu}$ appears (derivatives on $F$ can be replaced by derivatives
on $h$).  
In addition, the form of $S_{\rho_1\rho_2\mu_1\mu_2}$ is reminiscent
of the usual expression for the Riemann
tensor. Let us see if it is possible to make this analogy more
quantitative. The Riemann tensor, ${T^{\la}}_{\rho\mu\nu}$, for a non-symmetric metric  is given by (see
e.g. \cite{Damour:1993}; the ordering of
indices is important)
\be
{T^{\la}}_{\rho\mu\nu} = \pa_{\mu} {\Ups^{\la}}_{\rho\nu} -
{\Ups^{\si}}_{\rho\mu}{\Ups^{\la}}_{\si \nu} - (\mu \leftrightarrow \nu)\,,
\ee
where ${\Ups^{\la}}_{\rho\nu}$ is the connection. Let us further define
$\Ups_{\la\rho\nu} := h_{\la\si}{\Ups^{\si}}_{\rho\nu}$ and
$T_{\la\rho\mu\nu} := h_{\la\si}{T^{\si}}_{\rho\mu\nu}$. If we choose 
$\Ups_{\la\rho\nu} = \pa_{\rho}h_{\la\nu} = \pa_{\rho}F_{\la\nu}$, then a short calculation
leads to $T_{\la\rho\mu\nu} = S_{\la\rho\mu\nu}$. Thus, 
$S_{\la\rho\mu\nu}$ can be interpreted as the Riemann tensor for a
non-symmetric metric. 

The following useful relations for the derivatives of
$h_{\rho_1\rho_2}$ and $S_{\rho_1\rho_2\mu_1\mu_2}$ hold:
\bea \label{dS}
 \pa_{\mu} h_{\rho_1\rho_2} - h_{\rho_1
 \la}\Ups^{\la}{}_{\rho_2\mu}-\Ups_{\mu\rho_1\la}\de^{\la}_{\rho_2} =
 0\,, \non \\
\pa_{[\mu_3}S_{|\rho_1\rho_2|\mu_1\mu_2]} -
S_{\rho_1\la[\mu_1\mu_2}\Ups^{\la}{}_{|\rho_2|\mu_3]} -
\Ups_{[\mu_3|\rho_1\la}S^{\la}{}_{\rho_2|\mu_1\mu_2]} =0\,.
\eea
The first equation is analogous to the one showing that the
metric is covariantly constant for the case of a symmetric
metric, whereas the second one is similar to the Bianchi identity for
the usual Riemann tensor. 
Using the second equation in (\ref{dS}), it is straightforward to
check that 
$\D[\hat{\tr}(\mathsf{S}^{k})]=0$, thus showing that the corrections (\ref{Ws})
are all closed. 

Some further interesting relations are obtained upon 
dimensional reduction. We have seen in the previous section that
when the gauge field in the lower dimension is zero we obtain the
inverse of the induced metric from the part of $h^{MN}$ with both
indices longitudinal to the world volume. Similarly, from the
${}^{\rho}{}_{\mu \nu}$ part of $\Ups^{R}{}_{M N}$ we obtain 
${\Ga_{\!\!\mathrm{T}}}^{\la}_{\mu\nu}$, and
from the ${}^{i}{}_{\mu \nu}$ part we get  $-\Om^{i}_{\mu\nu}$ (the sign
can be removed by redefining
$X^i\rar-X^i$). To obtain these relations we used the static gauge.

So far we have discussed a second order formalism, i.e. one where the
connection is expressed in terms of the metric. There should also exist a 
first order formalism in which the Riemann tensor in the action is taken to depend on
the connection, and the relation between the connection and the metric
follows from the equation of motion for the connection. In the usual case one can always go between the two
formulations by varying the Einstein-Hilbert action in the
form $\int \sqrt{-g}\,g^{\mu\nu}R_{\mu\nu}(\Ga)$. In
our case it is not a priori clear which action to use as a replacement
for the this action. Notice for instance that one can  
construct more than one independent ``Ricci'' tensor, e.g. $T_{\rho_1\rho_2}:=h^{\mu_1\mu_2}T_{\rho_1\rho_2\mu_1\mu_2}$ and $\tilde{T}_{\rho_1\mu_1} =
h^{\rho_2\mu_2}T_{\rho_1\rho_2\mu_1\mu_2}$ (see \cite{Damour:1993} for
more details). Since the $\cO(\al')$ corrections vanish for
the superstring case we
get no clues from the string theory calculation. If the relation to
non-symmetric gravity is valid beyond supersymmetry then a calculation in
the bosonic string may lead to some insight. Unfortunately this
calculation has not been done to all orders in $F$.

We would like to stress that although the above results are very
suggestive it may be that the structure is only present for the terms
we have considered; it would therefore be interesting to determine
some other corrections, to see whether the structure persists. We
would also like to point out that the problems associated with theories with a
non-symmetric metric raised in \cite{Damour:1993} are not directly
applicable to our situation. In \cite{Damour:1993} only metrics of the
form $\de + B$ were discussed, which leads to problems with the
$\La$-symmetry, $B\rar B+\D \La$. Since we also have $F$ at our
disposal, these problems are circumvented. Also, we discuss
four-derivative corrections, whereas the focus in \cite{Damour:1993}
was on two-derivative terms. Theories with a non-symmetric metric have
also been dicussed within the context of D-brane effective actions in
\cite{Abou-Zeid:1997}. 

Is it possible to take the above discussion one step further and
somehow also
include a curved background metric and in this way incorporate the corrections involving the Riemann
curvature determined in \cite{Green:1997}? For the
nine-brane case one might try expressing everything in
terms of $h_{\mu\nu} = G_{\mu\nu} + F_{\mu\nu}$, where $G$ is the
background metric, and construct a 
$\Ups_{\la\rho\nu}$ which is such that it reduces to $\Ga_{\la\rho\nu}$
(possibly up to a multiplicative constant) when $h_{\mu\nu} =
G_{\mu\nu}$, so that $T_{\la\rho\mu\nu}$ reduces to
$R_{\la\rho\mu\nu}$. The answer appears to be no. It is possible to
find a $\Ups_{\la\rho\nu}$ which does the trick for 
the four-derivative corrections, but the same
expression will give erroneous results for the higher-form
corrections to the Wess-Zumino term. 

\section*{Acknowledgements}
The author would like to thank Michael Green and Kasper Peeters for discussions. This
work was supported by the European Commission under the contract
\hbox{FMBICT983302}. The author has also benefitted from the PPARC grant
PPA/G/S/1998/00613.

\section*{Appendices}

\appendix

\setcounter{equation}{0}
\section{Conventions} \label{convapp}
For convenience we have collected our conventions in this appendix. The
boundary of the disk is parameterised by the variable $\si$ which takes values
from 0 to $2\pi$. The string coordinate on the boundary is
$X^{\mu}(\si)$ and has the following expansion in terms of the $\al^{\mu}_{n}$ and $\tilde{\al}^{\mu}_{n}$ oscillators
\be 
X^{\mu}(\si) = x^{\mu} + i\sqrt{\ts{\frac{\al'}{2}}}\sum_{n\neq 0} \left(\frac{\al^{\mu}_{n}}{n}e^{in\si} + \frac{\tilde{\al}^{\mu}_{n}}{n}e^{-in\si}\right)\,.
\ee
The commutation relations are as usual
$[\al^{\mu}_m,\al^{\nu}_n]=m\,\eta^{\mu\nu}\de_{m+n}$ and
$[\tilde{\al}^{\mu}_m,\tilde{\al}^{\nu}_n]=m\,\eta^{\mu\nu}\de_{m+n}$,
from which it follows that $[X^{\mu}(\si),X^{\nu}(\xi)]=0$. The
boundary state $|B\rangle$ \cite{Callan:1987a,Callan:1988} satisfies (for the
nine-brane case) $P_{\mu} |B\rangle=0$ (where $P_{\mu}$ is the string
momentum). The explicit expression for
$|B\rangle$ is not needed in this paper;
it can be found e.g. in \cite{DiVecchia:1999a}. In the presence of a
gauge field $A_{\mu}(X)$, with field strength $F_{\mu\nu}(X) =
\frac{\pa}{\pa X^{\mu} } A_{\nu}(X) - \frac{\pa}{\pa X^{\nu}}A_{\mu}(X)$, the above condition is replaced by 
 $[2\pi\al'P_{\mu} + F_{\mu\nu}(X)\pa_{\si}X^{\nu}]|B(F(X))\rangle =
0$.  
Throughout the paper we absorb a factor of $2\pi \al'$ in
$F_{\mu\nu}$; what we call $F_{\mu\nu}$ is thus really
$2\pi\al'F_{\mu\nu}$. When we refer to derivative corrections we mean
corrections with additional factors of $\al'$. 

We use an unconventional normalisation for the fermionic
coordinates, $\Psi^{\mu}_{A}$, $A=-,+$. The canonical anti-commutation
relations between the fermions are taken to be
$\{ \Psi_{A}^{\mu}(\si),\Psi_{B}^{\nu}(\xi)\} =
-i\pi\al'\de^{\mu\nu}\de_{AB}\de(\si-\xi)$. The mode expansions are
\bea
&\mathrm{R}: &\left\{ \ba{rcl} \Psi^{\mu}_{-}(\si) &=& \sqrt{-\textstyle{\frac{i\al'}{2}}} \sum_{n\in\mathbb{Z} }d_{n}^{\mu}e^{in\si} \,,\non \\
   \Psi^{\mu}_{+}(\si) &=& \sqrt{-\textstyle{\frac{i\al'}{2}}}
   \sum_{n\in\mathbb{Z} }\tilde{d}_{n}^{\mu}e^{-in\si}  \,, \ea \right. \non \\
&\mathrm{NS}: &\left\{ \ba{rcl} \Psi^{\mu}_{-}(\si) &=& \sqrt{-\textstyle{\frac{i\al'}{2}}} \sum_{r\in\mathbb{Z}+\mbox{\tiny{$\frac{1}{2}$}} }b_{r}^{\mu}e^{ir\si}\,, \non \\
   \Psi^{\mu}_{+}(\si) &=& \sqrt{-\textstyle{\frac{i\al'}{2}}}
   \sum_{r\in\mathbb{Z}+\mbox{\tiny{$\frac{1}{2}$}}
   }\tilde{b}_{r}^{\mu}e^{-ir\si} \,, \ea \right.
\eea
with $\{d_m^{\mu},d^{\nu}_n\} = \de^{\mu\nu}\de_{m+n}$ and
$\{b_r^{\mu},b^{\nu}_s\} = \de^{\mu\nu}\de_{r+s}$ in the NS and R
sectors, respectively. 
We also use the notation $\Psi^{\mu} = \Psi^{\mu}_{-} + i \eta \Psi^{\mu}_{+}$ and
$\bar{\Psi}^{\mu} = \Psi^{\mu}_{-} - i \eta \Psi^{\mu}_{+}$.  The boundary state $|B\rangle$
satisfies $\bar{\Psi}^{\mu} |B\rangle=0$. In the presence of a gauge field
$A_{\mu}(X)$, the boundary condition reads $[\bar{\Psi}_{\mu}-
F_{\mu\nu}(X)\Psi^{\nu}]|B(F(X))\rangle = 0$.  To implement the GSO
projection  one
has to take certain linear combinations of the two spin-structure
sectors. The details are not important to us since the derivative
corrections are not sensitive to such considerations. 
   
We work in euclidian signature (our results can, however, easily be transformed to
Minkowski signature). Our normalisation of the boundary term in the $\si$-model is 
\be \label{simdl}
S = \frac{i}{2\pi\al'} \int d\si \left(\pa_{\si}X^{\mu}A_{\mu}(X) - \half
\Psi^{\mu}\Psi^{\nu} F_{\mu\nu}(X) \right) \,.
\ee
When convenient, we use superfields; our conventions are: $D = \tha
\pa_{\si} - \pa_{\tha}$, $\phi^{\mu} = X^{\mu} + \tha
\Psi^{\mu}$. In the superfield language the
boundary term can be written  $\frac{i}{2\pi\al'}\int d\si \D \tha
D_{\si}\phi^{\mu}A_{\mu}(\phi)$. 

Our conventions for differential forms are as follows: $X_p = \frac{1}{p!}
X_{\mu_1\cdots \mu_p} \D x^{\mu_1} \we \cdots \we \D x^{\mu_p}$ and
\be
X_p \we Y_q = \ts{\frac{1}{p!q!} X_{[\mu_1\cdots
\mu_p} Y_{\mu_{p+1} \cdots \mu_{p+q} ]}\D x^{\mu_1} \we \cdots \we \D x^{\mu_{p+q}} } \,.
\ee
We often suppress the wedge product. 
In the cases of branes with lower than maximal dimension we use the
following index conventions. Indices longitudinal to the brane (i.e. in
the world-volume directions) are denoted $\mu,\nu,\ldots$, indices in
the normal directions $i,j,\ldots$ and finally ten-dimensional bulk indices are labelled by
$M,N,\ldots$. For the nine-brane we use $\mu,\nu,\ldots$ to denote both
bulk and brane indices, since D9-branes are space filling. 

\setcounter{equation}{0}
\section{Technical details} \label{techapp}

In this appendix we will give some technical details of the calculations
underlying our results. Because of the complexity of the problem we
will focus on the main steps.

\subsection{Corrections to the Wess-Zumino term}
In this subsection we will discuss the method 
we used to derive the expressions for the corrections to the
Wess-Zumino term (\ref{2nform}), (\ref{Ws}) and the field
redefinitions needed to remove the two-form part at order
$\al'^2$. Here  
we will only consider the terms which have non-logarithmic finite
parts. This means that we exclude terms where two $\phi$'s within the
same factor are contracted, since $[G^{\mu\nu}_{12}]_{1\rar2}$ contains
a multiplicative $\ln \ep$ factor. 
In almost all expressions below we suppress an overall contraction of
$\langle C |\cdots \BF$. 

The first step of the calculations is to expand the exponential in (\ref{BFX}) using (\ref{Fexp}), insert the result into (\ref{WZBFX}) and collect terms of the
same order in derivatives. Next one uses the method outlined in section \ref{sectPre},
to rewrite the resulting expressions in terms of propagators
by calculating the various correlation functions. In this phase it is
imperative to use the fact that $F_{\mu\nu}$ satisfies the Bianchi identity 
\be
\label{Bianchi}
\pa_{[\rho}F_{\mu\nu]}=0
\Leftrightarrow \pa_{\rho}F_{\mu\nu} +
\pa_{\mu}F_{\nu\rho} + \pa_{\nu}F_{\rho \mu}
= 0 \,.
\ee
 Various symmetry properties of the propagator are also useful, such as
 $G_{12}^{\mu\nu} = G_{21}^{\nu\mu}$ and $D_{1}G^{\mu\nu}_{12} =
 (\tha_1-\tha_2)\pa_1 D_{12}^{\mu\nu}$. The latter result holds only
 in the R sector and follows
 from the fact that in this sector $K_{12}^{\mu\nu} =\pa_2
 D_{12}^{\mu\nu} = -\pa_1 D_{12}^{\mu\nu}$. 
Relabelling of indices and coordinates, as well as integration by parts
 can also profitably be used to relate various terms. Also, for the $\int \D\si
 \pa_{\si} \phi\,\phi^2 \pa F$ factors,  
the part with two fermionic zero modes vanishes since this term is a total derivative. 

Let us now discuss the $\al'^2$ corrections.  It is straightforward to
 check that the $\pa^3 F\pa F$ and $\pa^4 F$ contributions only involve
 logarithmic terms. Furthermore, the terms with no fermionic zero modes
 vanish by the general result given in section \ref{sectPre}.  Let us
 therefore start by discussing the
 terms with two zero modes.  For this case the $\pa^2 F \pa^2 F$ terms
 do
 not have any non-logarithmic finite parts. Examples of terms with non-logarithmic finite
 parts are:
\bea \label{2zm}
\rnode{y21}{\pa F} \, \rnode{y22}{\pa F} \, \rnode{y23}{\pa^{2} F}
\ncbar[nodesep=2pt,angle=-90,offsetA=4pt,offsetB=6pt,arm=2pt]{y21}{y22}
\ncbar[nodesep=2pt,angle=-90,offsetA=0pt,offsetB=2pt,arm=4pt]{y21}{y22}
\ncbar[nodesep=2pt,angle=-90,offsetA=-4pt,offsetB=4pt,arm=6pt]{y21}{y23}
\ncbar[nodesep=2pt,angle=-90,offsetA=2pt,offsetB=8pt,arm=2pt]{y22}{y23}
\psset{dotsize=2pt 0}
\ncbar[nodesep=2.8pt,angle=-90,offsetA=2pt,offsetB=-2pt,arm=0pt]{-*}{y23}{y23}
\ncbar[nodesep=2.8pt,angle=-90,offsetA=6pt,offsetB=-6pt,arm=0pt]{-*}{y23}{y23}
&=& \ts{\frac{1}{4}}(\ts{\frac{-i}{2\pi\al'}})^3\psi_0^{\eta} \psi_0^{\de} \int \D \si \pa_1
D_{13}^{\mu_1\la_1} D_{23}^{\mu_2\la_2} \pa_2 D_{12}^{\nu_1\nu_2}
D_{12}^{\rho_1 \rho_2} \pa_{\rho_1} F_{\mu_1\nu_1}\pa_{\rho_2} F_{\mu_2\nu_2} \pa_{\la_1}\pa_{\la_2}F_{\eta\de}\,,
\non \\
\rnode{y31}{\pa F} \, \rnode{y32}{\pa F} \, \rnode{y33}{\pa^{2} F}
\psset{dotsize=2pt 0}
\ncbar[nodesep=2.8pt,angle=-90,offsetA=6pt,offsetB=-6pt,arm=0pt]{-*}{y33}{y33}
\ncbar[nodesep=2.8pt,angle=-90,offsetA=0pt,offsetB=-0pt,arm=0pt]{-*}{y32}{y32}
\ncbar[nodesep=2pt,angle=-90,offsetA=2pt,offsetB=4pt,arm=2pt]{y31}{y32}
\ncbar[nodesep=2pt,angle=-90,offsetA=4pt,offsetB=6pt,arm=2pt]{y32}{y33}
\ncbar[nodesep=2pt,angle=-90,offsetA=-6pt,offsetB=-2pt,arm=6pt]{y31}{y33}
\ncbar[nodesep=2pt,angle=-90,offsetA=-2pt,offsetB=2pt,arm=4pt]{y31}{y33}
 &=& (\ts{\frac{-i}{2\pi\al'}})^3 \psi_0^{\nu_2} \psi_0^{\de} \int \D \si \Big[ K_{23}^{\mu_2\eta}K_{13}^{\nu_1\la_1}
 D_{12}^{\mu_1\rho_2} D_{13}^{\rho_1\la_2}
 \non \\ && \qquad \qquad \qquad +\, K_{12}^{\mu_1\mu_2}
 K_{13}^{\nu_1\eta}
 D_{13}^{\rho_1\la_1}D_{23}^{\rho_2\la_2}\Big]\pa_{\rho_1}
 F_{\mu_1\nu_1}\pa_{\rho_2} F_{\mu_2\nu_2}
 \pa_{\la_1}\pa_{\la_2}F_{\eta\de} \,.
\eea
In addition, there are also two different types of contractions of the  
$\rnode{v11}{\pa F} \, \rnode{v12}{\pa F} \,\rnode{v13}{\pa F} \, \rnode{v14}{\pa F}
\psset{dotsize=2pt 0}
\ncbar[nodesep=2.8pt,angle=-90,offsetA=0pt,offsetB=-0pt,arm=0pt]{-*}{v11}{v11}
\ncbar[nodesep=2.8pt,angle=-90,offsetA=0pt,offsetB=-0pt,arm=0pt]{-*}{v12}{v12}
\ncbar[nodesep=2.8pt,angle=-90,offsetA=0pt,offsetB=-0pt,arm=0pt]{-*}{v13}{v13}
\ncbar[nodesep=2.8pt,angle=-90,offsetA=0pt,offsetB=-0pt,arm=0pt]{-*}{v14}{v14}$
 terms with non-logarithmic finite parts.
A few words about our notation are in order. The dots in the
above expressions indicate zero modes. Furthermore, $\D\si$ is
short-hand notation for  
 $\D\si_1 \cdots \D\si_k$, where $k$ equals the number of $F$'s. All integrations
 range from $0$ to $2\pi$. The contractions
 and the placement of dots are schematical. For instance, 
 a line that seems to end on an $F$ is not necessarily a contraction with
 an index on that $F$, it could also be a contraction with an index on
a partial derivative acting on that $F$. 

To perform the integration over the $\si$'s, one uses $\int \D\si e^{i n \si} =
2\pi \de_{n,0}$. 
As an example, for the second expression in (\ref{2zm}) one gets
\bea \label{dFdFd2F}
\rnode{z31}{\pa F} \, \rnode{z32}{\pa F} \, \rnode{z33}{\pa^{2} F}
\psset{dotsize=2pt 0}
\ncbar[nodesep=2.8pt,angle=-90,offsetA=6pt,offsetB=-6pt,arm=0pt]{-*}{z33}{z33}
\ncbar[nodesep=2.8pt,angle=-90,offsetA=0pt,offsetB=-0pt,arm=0pt]{-*}{z32}{z32}
\ncbar[nodesep=2pt,angle=-90,offsetA=2pt,offsetB=4pt,arm=2pt]{z31}{z32}
\ncbar[nodesep=2pt,angle=-90,offsetA=4pt,offsetB=6pt,arm=2pt]{z32}{z33}
\ncbar[nodesep=2pt,angle=-90,offsetA=-6pt,offsetB=-2pt,arm=6pt]{z31}{z33}
\ncbar[nodesep=2pt,angle=-90,offsetA=-2pt,offsetB=2pt,arm=4pt]{z31}{z33}
 &=& -i\al' \psi_0^{\nu_2} \psi_0^{\de}
 \ds{ \sum_{n,p=1}^{\infty} } \ts{\frac{e^{-\ep(3p{+}2m)}}{p(m{+}p)} }
 \Big[
 h^{\la_2\rho_1}h^{\mu_1\rho_2}h^{\mu_2\eta}h^{\nu_1\la_1}+h^{\rho_1\la_2}h^{\rho_2\mu_1}h^{\eta\mu_2}h^{\la_1\nu_1} \non \\ && \qquad \qquad\qquad\qquad\quad\;\;\; -\,h^{\eta\nu_1}h^{\mu_2\mu_1}h^{\la_2\rho_2}h^{\rho_1\la_1}-h^{\nu_1\eta}h^{\mu_1\mu_2}h^{\rho_2\la_2}h^{\la_1\rho_1}\Big] \non \\ && \qquad \qquad\qquad\qquad\quad\;\;\;\times\pa_{\rho_1}
 F_{\mu_1\nu_1}\pa_{\rho_2} F_{\mu_2\nu_2}
 \pa_{\la_1}\pa_{\la_2}F_{\eta\de}\,.
\eea
The  sums arising from the integrations are evaluated in appendix
\ref{sumapp}. Since it is known that $\langle C  \BF \rar T_9
Ce^{F}$, and that only the zero mode part of $\BF$ contributes to this result, one can
read off the rule $\langle C| (\psi_0^{\mu}\psi_0^{\nu}X_{\mu\nu})^{k}
\BF \rar T_9 (-2i\al')^k C e^F X^{k}$. This rule implies that the explicit form of $|C \rangle
$ is not needed in this paper and also gives the relation
between the number of zero modes and the form degree of the resulting
correction. It furthermore shows that the above two-form expressions, which are all of the form
$\psi_0^{\mu}\psi_0^{\nu}X_{\mu\nu}$, can be removed by the field
redefinition $\de F = 2i\al'X$, which is an expression of order
$\al'^2$. Whether this can in turn be interpreted as a redefinition of
$A_{\mu}$ depends on the renormalisation scheme but there has to exist
a scheme in which it is possible; see section
\ref{sectPre} for further discussion of this point. 

The $\al'^2$ four-form four-derivative corrections are easier to calculate than the two-form four-derivative
ones, since 
there are more zero-modes and consequently fewer
propagators. In fact, without too much additional effort the
corresponding $2n$-form
$2n$-derivative corrections can also be determined. The divergent parts
of these expressions are removed by the terms induced by the first
order field redefinition (\ref{deF1}). We will not give the details
here. Instead, we will perform one sample calculation of the finite
part for one particular 6-form
6-derivative correction, namely the $(\pa F)^4\pa^2
F$ term. The calculation proceeds as follows: 
\bea \label{6zm}
\rnode{z41}{\pa F} \, \rnode{z42}{\pa F} \,  \rnode{z43}{\pa F} \,
\rnode{z44}{\pa F} \, \rnode{z45}{\pa^{2} F}
\psset{dotsize=2pt 0}
\ncbar[nodesep=2.8pt,angle=-90,offsetA=6pt,offsetB=-6pt,arm=0pt]{-*}{z45}{z45}
\ncbar[nodesep=2.8pt,angle=-90,offsetA=2pt,offsetB=-2pt,arm=0pt]{-*}{z45}{z45}
\ncbar[nodesep=2.8pt,angle=-90,offsetA=0pt,offsetB=-0pt,arm=0pt]{-*}{z41}{z41}
\ncbar[nodesep=2.8pt,angle=-90,offsetA=0pt,offsetB=-0pt,arm=0pt]{-*}{z42}{z42}
\ncbar[nodesep=2.8pt,angle=-90,offsetA=0pt,offsetB=-0pt,arm=0pt]{-*}{z43}{z43}
\ncbar[nodesep=2.8pt,angle=-90,offsetA=0pt,offsetB=-0pt,arm=0pt]{-*}{z44}{z44}
\ncbar[nodesep=2pt,angle=-90,offsetA=4pt,offsetB=4pt,arm=2pt]{z41}{z42}
\ncbar[nodesep=2pt,angle=-90,offsetA=4pt,offsetB=4pt,arm=2pt]{z42}{z43}
\ncbar[nodesep=2pt,angle=-90,offsetA=4pt,offsetB=4pt,arm=2pt]{z43}{z44}
\ncbar[nodesep=2pt,angle=-90,offsetA=-4pt,offsetB=4pt,arm=4pt]{z41}{z45}
\ncbar[nodesep=2pt,angle=-90,offsetA=4pt,offsetB=8pt,arm=2pt]{z44}{z45}
 &=& \ts{\frac{5}{5!}} (\ts{\frac{-i}{2\pi\al'}})^5 \ts{\frac{1}{2^2}
 } \non \\[4pt] && \hspace{-3cm}\times \ts{\int} \D \si \langle
 C|(\psi_0^{\nu_1} \psi_0^{\nu_2} \psi_0^{\nu_3} \psi_0^{\nu_4}
 \psi_0^{\eta}\psi_0^{\de})\Psi^{\mu_1}X^{\rho_1}\Psi^{\mu_2}X^{\rho_2}\Psi^{\mu_3}X^{\rho_3}\Psi^{\mu_4}X^{\rho_4}X^{\la_1}X^{\la_2} \BF \non \\ && \times \, \pa_{\rho_1} F_{\mu_1\nu_1}\pa_{\rho_2} F_{\mu_2\nu_2}\pa_{\rho_3} F_{\mu_3\nu_3}\pa_{\rho_4} F_{\mu_4\nu_4}\pa_{\la_1\la_2} F_{\eta\de} \non \\ &=& \ts{\frac{1}{2^2}} (\ts{\frac{-i}{2\pi\al'}})^5 (\psi_0)^6\int \D \si [K_{12}^{\mu_1\mu_2} K_{34}^{\mu_3\mu_4} D_{15}^{\rho_1\la_1} D_{35}^{\rho_3\la_2} D_{24}^{\rho_2\rho_4}] (\pa F)^4 \pa^2 F \non \\ &=& \ts{\frac{1}{2^2}} (\ts{\frac{-i}{2\pi\al'}})^5
 (2\pi\al')^5(\psi_0)^6\Big[ h^{\mu_1\mu_2} h^{\mu_4\mu_3}
 h^{\la_1\rho_1} h^{\rho_3\la_2} h^{\rho_2\rho_4} \non \\ && + \,h^{\mu_2\mu_1} h^{\mu_3\mu_4}
 h^{\rho_1\la_1} h^{\la_2\rho_3} h^{\rho_4\rho_2} \Big]
 \sum_{n=1}\ts{\frac{e^{-5\ep n}}{n^3}} (\pa F)^4\pa^2F\,.
\eea
We would like to stress that we only kept the terms with
non-logarithmic finite parts, which is why we discarded some of the
possible contractions. We have also suppressed the
$\langle C|\cdots\BF$ contraction in the last two expressions.  
By relabelling of indices one can easily show that in the last
expression in (\ref{6zm}) the two $h^5$ terms 
are equal. Finally, removing the regulator and 
using the rule $\langle C| (\psi_0^{\mu}\psi_0^{\nu}X_{\mu\nu})^{k}
\BF \rar T_9 (-2i\al')^k C e^F X^{k}$, reproduces the $(\pa F)^4\pa^2
F$ part of $W_6$ given in (\ref{Ws}).

\subsection{Corrections to the Born-Infeld term}
In this subsection we will give some details of the calculation of the
$\al'^2$ four-derivative corrections to the Born-Infeld action. 
The first step is to expand the exponential in  (\ref{BFX}),
(\ref{BIBFX}) using (\ref{Fexp}) and collect terms of the
same order in derivatives and rewrite the expressions in terms of propagators
by calculating the various correlation functions (or, equivalently, by
writing down the various Feynman diagrams). The expressions greatly
simplify as a consequence of the Bianchi
identity, $\pa_{[\rho}F_{\mu\nu]}=0$. At this stage of the calculation various properties of the propagator are also useful, such as 
\bea
\label{Gprop}
&&G_{12}^{\mu\nu} = G_{21}^{\nu\mu}\,, \non \\ 
&&\ts{\int} \D \si_1 G_{12}^{\mu\nu} =0\,, \non \\
&&D_{1}G_{12}^{\mu\nu} = - D_{2}G_{12}^{\mu\nu}\,, \non \\ 
&&D_{1}G_{12}^{\mu\nu}D_{2}G_{12}^{\eta\de}=
D_{1}G_{12}^{\eta\de}D_{2}G_{12}^{\mu\nu}\,.
\eea 
The second condition above implies that (in a diagrammatical approach)
only diagrams (not neccessarily connected) whose parts are
one-particle irreducible contribute. 
Relabelling of indices and integrations by parts also leads to a
reduction of the number of independent terms. 

The $\al'^2$ corrections can be classified according to how the index contractions between the
different $\pa^n F$'s are performed. The contributions with
non-logarithmic finite parts become (after simplifications) 
\bea \label{al2prop}
\rnode{x21}{\pa^{2} F} \, \rnode{x22}{\pa^{2} F}
\ncbar[nodesep=2pt,angle=-90,offsetA=-8pt,offsetB=4pt,arm=2pt]{x21}{x21}
\aput[-7pt]{:U}{\mbox{\tiny{$DG$}}}
\ncbar[nodesep=2pt,angle=-90,offsetA=3pt,offsetB=-7pt,arm=2pt]{x22}{x22}
\aput[-7pt]{:U}{\mbox{\tiny{$DG$}}}
\ncbar[nodesep=2pt,angle=-90,offsetA=6pt,offsetB=6pt,arm=2pt]{x21}{x22}
\ncbar[nodesep=2pt,angle=-90,offsetA=2pt,offsetB=2pt,arm=4pt]{x21}{x22}
 &=&
 \ts{\frac{1}{2!}}(\ts{\frac{-i}{2\pi\al'}})^2\ts{\frac{1}{8^2}}(-8)\int \D \si_1 \D\si_2 \D \tha_1 \D \tha_2
\Big[ [D_1 G_{12}^{\nu\mu}]_{2\rar 1} [D_2 G_{12}^{\de\eta}]_{1\rar2}
G_{12}^{\rho_1\la_1}G_{12}^{\rho_2\la_2}\Big] \non \\ &&\qquad\qquad \qquad\times\,
\pa_{\rho_1}\pa_{\rho_2}F_{\mu\nu} \pa_{\la_1}\pa_{\la_2}F_{\eta\de} \non \\
\rnode{x11}{\pa^2 F} \, \rnode{x12}{\pa^2 F} 
\ncbar[nodesep=2pt,angle=-90,offsetA=0pt,offsetB=4pt,arm=4pt]{x11}{x12} 
\ncbar[nodesep=2pt,angle=-90,offsetA=-4pt,offsetB=0pt,arm=6pt]{x11}{x12} 
\ncbar[nodesep=2pt,angle=-90,offsetA=4pt,offsetB=8pt,arm=2pt]{x11}{x12} 
\ncbar[nodesep=2pt,angle=-90,offsetA=-8pt,offsetB=-4pt,arm=8pt]{x11}{x12}
&=& \ts{\frac{1}{2!}}(\ts{\frac{-i}{2\pi\al'}})^2\ts{\frac{1}{8^2}}(-16)\int \D \si_1 \D\si_2 \D \tha_1 \D \tha_2 \Big[ D_1
G_{12}^{\nu\eta} D_2 G_{12}^{\mu\de}
G_{12}^{\rho_1\la_1}G_{12}^{\rho_2\la_2}\Big] \non \\  && \qquad \qquad \qquad\times \,
\pa_{\rho_1}\pa_{\rho_2}F_{\mu\nu} \pa_{\la_1}\pa_{\la_2}F_{\eta\de}
\non \\
\rnode{x31}{\pa F}\,\rnode{x32}{\pa F} \, \rnode{x33}{\pa^{2} F}
\ncbar[nodesep=2pt,angle=-90,offsetA=3pt,offsetB=-7pt,arm=2pt]{x33}{x33}
\aput[-7pt]{:U}{\mbox{\tiny{$DG$}}} 
\ncbar[nodesep=2pt,angle=-90,offsetA=4pt,offsetB=6pt,arm=2pt]{x31}{x32}
\ncbar[nodesep=2pt,angle=-90,offsetA=0pt,offsetB=2pt,arm=4pt]{x31}{x32}
\ncbar[nodesep=2pt,angle=-90,offsetA=-4pt,offsetB=4pt,arm=6pt]{x31}{x33}
\ncbar[nodesep=2pt,angle=-90,offsetA=2pt,offsetB=8pt,arm=2pt]{x32}{x33}
 &=&
\ts{\frac{3}{3!}}(\ts{\frac{-i}{2\pi\al'}})^3\ts{\frac{1}{3^2\!\cdot 8}}(-36)\int \D \si \D \tha \Big[ [D_3 G_{23}^{\de\eta}]_{2\rar3} \Big({-}D_{1}G_{12}^{\nu_1\nu_2}
D_{2}G_{23}^{\mu_2\la_2} G_{13}^{\mu_1\la_1} G_{12}^{\rho_1\rho_2}
\non \\ && \qquad +\,
\half D_{1}G_{12}^{\nu_1\mu_2}
D_{2}G_{12}^{\mu_1\nu_2} G_{13}^{\rho_1\la_1}
G_{23}^{\rho_2\la_2}\Big)\Big] \pa_{\rho_1}F_{\mu_1\nu_1}
\pa_{\rho_2}F_{\mu_2\nu_2} \pa_{\la_1}\pa_{\la_2}F_{\eta\de} \non \\
\rnode{x41}{\pa F}\,\rnode{x42}{\pa F} \, \rnode{x43}{\pa^{2} F}
\ncbar[nodesep=2pt,angle=-90,offsetA=4pt,offsetB=4pt,arm=2pt]{x41}{x42}
\ncbar[nodesep=2pt,angle=-90,offsetA=0pt,offsetB=0pt,arm=6pt]{x41}{x43}
\ncbar[nodesep=2pt,angle=-90,offsetA=-4pt,offsetB=-4pt,arm=8pt]{x41}{x43}
\ncbar[nodesep=2pt,angle=-90,offsetA=4pt,offsetB=8pt,arm=2pt]{x42}{x43}
\ncbar[nodesep=2pt,angle=-90,offsetA=0pt,offsetB=4pt,arm=4pt]{x42}{x43}
 &=&
\ts{\frac{3}{3!}}(\ts{\frac{-i}{2\pi\al'}})^3\ts{\frac{1}{3^2\!\cdot
8}}(-72)\int \D \si \D \tha \Big[ D_{1}G_{13}^{\nu_1\eta}D_{2}G_{23}^{\nu_2\la_1} D_{3}G_{13}^{\mu_1\de}
 G_{12}^{\rho_1\mu_2}
G_{23}^{\rho_2\la_2} \non \\ && \quad + \,
 D_{1}G_{12}^{\nu_1\mu_2}D_{2}G_{23}^{\nu_2\eta}D_{3}G_{13}^{\mu_1\de}
 G_{13}^{\rho_1\la_1}
G_{23}^{\rho_2\la_2} \Big]  \pa_{\rho_1}F_{\mu_1\nu_1}
\pa_{\rho_2}F_{\mu_2\nu_2} \pa_{\la_1}\pa_{\la_2}F_{\eta\de}  \non \\
\rnode{x51}{\pa F}\,\rnode{x52}{\pa F} \, \rnode{x53}{\pa F} \,
\rnode{x54}{\pa F}
\ncbar[nodesep=2pt,angle=-90,offsetA=4pt,offsetB=4pt,arm=2pt]{x51}{x52}
\ncbar[nodesep=2pt,angle=-90,offsetA=0pt,offsetB=0pt,arm=4pt]{x51}{x52}
\ncbar[nodesep=2pt,angle=-90,offsetA=4pt,offsetB=4pt,arm=2pt]{x53}{x54}
\ncbar[nodesep=2pt,angle=-90,offsetA=0pt,offsetB=0pt,arm=4pt]{x53}{x54}
\ncbar[nodesep=2pt,angle=-90,offsetA=4pt,offsetB=4pt,arm=2pt]{x52}{x53}
\ncbar[nodesep=2pt,angle=-90,offsetA=-4pt,offsetB=-4pt,arm=6pt]{x51}{x54}
&=&
\ts{\frac{1}{4!}}(\ts{\frac{-i}{2\pi\al'}})^4\ts{\frac{3^5}{3^4}}\int
\D \si \D \tha \Big\{ \half D_1
G_{12}^{\nu_1\mu_2} D_2 G_{12}^{\mu_1\nu_2} D_3 G_{34}^{\nu_3\mu_4} D_4
G_{34}^{\mu_3\nu_4} G_{13}^{\rho_1\rho_3} G_{24}^{\rho_2\rho_4} \non
\\ && \quad +\, D_1G_{12}^{\nu_1\mu_2}D_2G_{23}^{\nu_2\mu_3}D_3
G_{34}^{\nu_3\mu_4}D_4G_{14}^{\mu_1\nu_4}G_{12}^{\rho_1\rho_2}G_{34}^{\rho_3\rho_4}
\non \\ &&\quad +\,
\half[D_{1}G_{12}^{\nu_1\mu_2}D_2G_{12}^{\mu_1\nu_2}D_3G_{13}^{\rho_1\nu_3}D_4G_{34}^{\mu_3\nu_4}G_{24}^{\rho_2\mu_4}G_{34}^{\rho_3\rho_4}
+ (\mu_4\leftrightarrow \rho_4)] \non \\ && \quad +\,
\half[D_{1}G_{12}^{\nu_1\mu_2}D_2G_{12}^{\mu_1\nu_2}D_3G_{13}^{\rho_1\nu_3}D_4G_{34}^{\mu_3\nu_4}G_{24}^{\rho_2\mu_4}G_{34}^{\rho_3\rho_4}
+ (\mu_3\leftrightarrow \rho_3)] \Big\} \non \\ && \qquad\qquad \qquad
\qquad\times \,\pa_{\rho_1} F_{\mu_1\nu_1}\pa_{\rho_2}
F_{\mu_2\nu_2}\pa_{\rho_3}F_{\mu_3\nu_3}\pa_{\rho_4}F_{\mu_4\nu_4} \non \\ 
\rnode{x61}{\pa F}\,\rnode{x62}{\pa F} \, \rnode{x63}{\pa F} \,
\rnode{x64}{\pa F}
\ncbar[nodesep=2pt,angle=-90,offsetA=4pt,offsetB=4pt,arm=2pt]{x61}{x62}
\ncbar[nodesep=2pt,angle=-90,offsetA=0pt,offsetB=0pt,arm=4pt]{x61}{x63}
\ncbar[nodesep=2pt,angle=-90,offsetA=4pt,offsetB=4pt,arm=2pt]{x63}{x64}
\ncbar[nodesep=2pt,angle=-90,offsetA=-4pt,offsetB=0pt,arm=6pt]{x61}{x64}
\ncbar[nodesep=2pt,angle=-90,offsetA=4pt,offsetB=4pt,arm=2pt]{x62}{x63}
\ncbar[nodesep=2pt,angle=-90,offsetA=0pt,offsetB=-4pt,arm=8pt]{x62}{x64}
 &=& \ts{\frac{1}{4!}}(\ts{\frac{-i}{2\pi\al'}})^4\ts{\frac{3^4}{3^4}} \int \D \si \D \tha \Big\{ D_{1}G_{12}^{\nu_1\mu_2}
D_{2}G_{23}^{\nu_2\mu_3}D_{3}G_{34}^{\nu_3\mu_4}\Big[3D_{4}G_{14}^{\mu_1\nu_4} G_{13}^{\rho_1\rho_3}
G_{24}^{\rho_2\rho_4} \non \\ && \qquad\qquad\qquad\qquad +\, 2D_{4}G_{24}^{\rho_2\nu_4} ( G_{14}^{\rho_1\rho_4}
G_{23}^{\mu_1\rho_3} + G_{14}^{\mu_1\rho_4}
G_{23}^{\rho_1\rho_3}) \Big] \Big\} \\ \non && \qquad\qquad \qquad \qquad\times \,\pa_{\rho_1} F_{\mu_1\nu_1}\pa_{\rho_2} F_{\mu_2\nu_2}\pa_{\rho_3}F_{\mu_3\nu_3}\pa_{\rho_4}F_{\mu_4\nu_4}
\eea
In the fomul\ae{} above we have suppressed an overall factor $\langle 0 \BF$,
which will give an overall factor $-T_9\sqrt{\det\,h}$ in the
correction to the action. Also, the contraction symbols are
schematical; there is no distinction between lines ending on an $F$
and on a derivative acting on that $F$. The notation `$DG$' means that
the contractions are between a $D\phi$ and a $\phi$. 

The next step is to perform the integrations over the $\si$'s and
$\tha$'s. We have written the terms in (\ref{al2prop}) so that all $D$'s act on
different $G^{\mu\nu}$'s. This way of writing the expressions makes it very
easy to incorporate the fermionic modes. Since there are as many
$D$'s as there are $\D \tha$'s in the above expressions and $D_1 G_{12}^{\mu\nu} = \tha_1 \pa_1
D_{12}^{\mu\nu} + \tha_2 K_{12}^{\mu\nu}$, the integrations
over the $\tha$'s 
pick out one of the terms (either a $\pa D$ or a $K$) in each of the
$DG$'s (the $G$'s without derivatives get
replaced by $D$'s). Furthermore, since
$K_{12}^{\mu\nu}$ and $\pa_2 D_{12}^{\mu\nu} = - \pa_1
D_{12}^{\mu\nu}$ are equal after the replacement of integer modes with
half-integer ones, the resulting expressions where the integration
over the 
$\tha$'s picks out some of the $K$'s instead of the $\pa D$'s, is
obtained by simply changing the relevant summation indices from sums
over integers to sums over half integers (sometimes the overall sign also
changes). 
Notice that not all combinations of $\pa D$'s and $K$'s are
allowed in all instances.  The integrations over the $\si$'s will lead to a collection
of Kronecker $\de$'s (using $\int_{0}^{2 \pi} \D \si e^{i n \si} = 2 \pi
\de_{n,0}$).  

The final step is to evaluate the resulting sums and extract the finite
 parts which will give the contribution to effective action.  
 The power divergent terms cancel between the bosonic and
fermionic sectors as a result of supersymmetry. Since the theory is
renormalisable the logarithmically divergent terms can be removed by
absorbing them in $A_{\mu}$, although we have not verified this fact in
 detail (the results in \cite{Andreev:1988b} should prove to be useful
when addressing this question). For the finite terms an additional
complication arises: some
 of these terms are renormalisation scheme dependent, and one would
 like to get rid of these terms as well. As we have seen in section
 \ref{sectPre} it is not
 possible to remove all the renormalisation scheme dependent terms within
 the scheme used in this paper. Instead another renormalisation scheme
 has to be used, 
 such as the scheme discussed in
 \cite{Andreev:1988}. Fortunately, there is a way around this problem
 as we will see below.  

As a sample calculation let us discuss the third type of terms in
(\ref{al2prop}). Let us start with the first term. We
first note that $[D_3 G_{23}^{\de\eta}]_{2\rar3}
= \tha_3(-2i\al')h_{A}^{\de\eta}[\sum_n e^{-\ep n}{-}\sum_r e^{-\ep r}]$,
which, when the regulator is removed, becomes $\tha_3(-\frac{1}{2})(-2i\al')h_{A}^{\de\eta}$. Furthermore, the
 integrations over $\tha_1$, $\tha_2$ picks out
the  $-\pa_{1}D_{12}^{\nu_1\nu_2}
\pa_{2}D_{23}^{\mu_2\la_2}$ part from $D_{1}G_{12}^{\nu_1\nu_2}
D_{2}G_{23}^{\mu_2\la_2}$. We now note that under a
field redefinition, the lowest order Born-Infeld term transforms as
$\de\sqrt{\det(1+F)} = -\frac{1}{2}h^{\mu\nu}\de F_{\mu\nu}$. The term
induced by the field redefinition required to remove the first term in (\ref{2zm}) thus has exactly the right
form to cancel the contribution under discussion. After some
relabelling of indices and use of symmetry properties of the
propagator, we see that this is indeed the case. Thus, the vanishing of the
 $\psi_0^2$ contribution to the Wess-Zumino term induces terms which
simplify the corrections to the Born-Infeld part as well. For the
present term the cancellation is independent of the renormalisation
scheme used, but in general the  vanishing of the
 $\psi_0^2$ contribution to the Wess-Zumino term leads to a strong
 indication as to which redefinition of
 $F_{\mu\nu}$ is needed to remove the scheme dependent terms. Some of
 the terms that are removed in this way involve e.g.~$\Li_2(-2)$ (see appendix \ref{sumapp} for properties of
 dilogarithms), which can not be reexpressed in terms of familiar
 constants. 

We now turn to the
second term in the third type of terms in
(\ref{al2prop}). We get 
\bea 
&&\ts{\frac{3}{3!}}(\ts{\frac{-i}{2\pi\al'}})^3\ts{\frac{1}{3^2\!\cdot 8}}(-36)(\half)\int \D \si \D \tha [D_3 G_{23}^{\de\eta}]_{2\rar3} 
D_{1}G_{12}^{\nu_1\mu_2}
D_{2}G_{12}^{\mu_1\nu_2} G_{13}^{\rho_1\la_1}
G_{23}^{\rho_2\la_2} \non \\ && \qquad\qquad\qquad\qquad\times \,\pa_{\rho_1}F_{\mu_1\nu_1}
\pa_{\rho_2}F_{\mu_2\nu_2} \pa_{\la_1}\pa_{\la_2}F_{\eta\de} \non \\
&=& \ts{\frac{1}{8}}(\ts{\frac{-i}{2\pi\al'}})^3
\al'h_{A}^{\de\eta}(2\pi\al')^3\al'\Big\{4 h^{\mu_1\nu_2}h^{\mu_2\nu_1}h^{\rho_1\la_1}h^{\la_2\rho_2}[\sum_{n,p=1}^{\infty}\ts{\frac{e^{-\ep(3p+2n)}}{p^2}}
- (n\leftrightarrow r)] \non \\ && \qquad-\,
2h^{\mu_2\nu_1}h^{\mu_2\nu_1}h^{\rho_1\la_1}h^{\la_2\rho_2}[\sum_{m,n=1}^{\infty}\ts{\frac{e^{-2\ep(m+n)}}{(m+n)^2}}
- (n,m\leftrightarrow r,s)]\Big\} \pa F\pa F \pa^2 F\non \\ &=& \al'^2 \ts{\frac{\pi^2}{12}}h_{A}^{\eta\de}h_{A}^{\mu_2\mu_1}h^{\nu_2\nu_1}h^{\rho_1\la_1}h^{\la_2\rho_2} \pa_{\rho_1}F_{\mu_1\nu_1}
\pa_{\rho_2}F_{\mu_2\nu_2} \pa_{\la_1}\pa_{\la_2}F_{\eta\de}\,,
\eea
where we have used results listed in appendix \ref{sumapp} to evaluate
the sums. Again, we have only kept the terms with non-logarithmic finite parts. We see
that the above result reproduces the $\pa F \pa F \pa^2 F$ part of 
$h^{\rho_4\rho_1}h^{\rho_4\rho_1}S_{\rho_1\rho_2}S_{\rho_3\rho_4}$
 in (\ref{BI2}). 

The story is similar for the other terms in (\ref{al2prop}). The
$\pa^2F\pa^2 F$ terms give the corresponding contributions to
$h^{\rho_4\rho_1}h^{\rho_4\rho_1}S_{\rho_1\rho_2}S_{\rho_3\rho_4}$ and
$h^{\rho_4\rho_1}h^{\rho_4\rho_1}h^{\mu_4\mu_1}h^{\mu_4\mu_1}S_{\rho_1\rho_2\mu_1\mu_2}S_{\rho_3\rho_4\mu_3\mu_4}$.
For the case of the remaining type of $\pa F \pa F \pa^2 F$ terms, the first
part is removed by the term induced by the field redefinition
needed to cancel (\ref{dFdFd2F}), whereas the second part gives the $\pa F \pa F \pa^2
F$ part of
$h^{\rho_4\rho_1}h^{\rho_4\rho_1}h^{\mu_4\mu_1}h^{\mu_4\mu_1}S_{\rho_1\rho_2\mu_1\mu_2}S_{\rho_3\rho_4\mu_3\mu_4}$.
For the first type of $(\pa F)^4$ terms in (\ref{al2prop}), the first
term gives the corresponding part of
$h^{\rho_4\rho_1}h^{\rho_4\rho_1}S_{\rho_1\rho_2}S_{\rho_3\rho_4}$, whereas the second term gives rise to half of the $(\pa F)^4$ part of
$h^{\rho_4\rho_1}h^{\rho_4\rho_1}h^{\mu_4\mu_1}h^{\mu_4\mu_1}S_{\rho_1\rho_2\mu_1\mu_2}S_{\rho_3\rho_4\mu_3\mu_4}$.
The other terms are cancelled by terms induced by field redefinitions
needed to remove two-form $(\pa F)^4$ contributions to the Wess-Zumino
term. Finally, for the second type of $(\pa F)^4$ terms, the first term
gives rise to the rest of $h^{\rho_4\rho_1}h^{\rho_4\rho_1}h^{\mu_4\mu_1}h^{\mu_4\mu_1}S_{\rho_1\rho_2\mu_1\mu_2}S_{\rho_3\rho_4\mu_3\mu_4}$,
whereas the other terms are removed by field redefinitions.

In conclusion, we would like to stress that given the large
 number of contributions involved, the final result is surprisingly simple.

\subsection{Four-derivative four-$F$ terms}
In this subsection we will show that when restricting the expression
in (\ref{BI2}) to the terms with at most four $F$'s, we correctly
reproduce the previously known result (\ref{t8F4}). Firstly, let us
note that any term involving four $\pa$'s and two $F$'s can be removed
by a field redefinition since by using the Bianchi identity and
integrations by parts such a term is necessarily proportional to the
lowest order equation of motion, $\pa^{\mu}F_{\mu\nu}$=0. Also, the $F^4$ terms induced by the field redefinition removing the $F^2$ terms can again be removed. Secondly, any term of the form $FF\pa^2F \pa^2F$ can be rewritten as $F \pa F \pa F \pa^2 F$ up to terms which are proportional to the lowest order equation of motion, by using the Bianchi identity and integration by parts. To show this fact we note that at least four indices have to be contracted within the $\pa^2 F \pa^2 F$ part. 
Furthermore, these contractions have to be between the two factors
since otherwise we get terms proportional to the equation of
motion. If one of the contracted indices is on a derivative we
integrate by parts with repect to this derivative, which will give us
$F \pa F \pa F \pa^2 F$ terms modulo terms proportional to the
equation of motion. The only remaining possibility is $F^{\mu\nu}
F^{\eta \de} \pa_{\mu}\pa_{\eta} F_{\rho\la} \pa_{\nu}\pa_{\de}
F^{\rho\la}$, which by integration with  respect to $\pa_{\mu}$ gives $F \pa F \pa F \pa^2 F$ terms. Finally, any $(\pa F)^4$ can, by integration by parts, be rewritten in terms of $F \pa F \pa F \pa^2 F$ terms. We thus arrive at the conclusion that any term with four $F$'s and four $\pa$'s can (modulo terms proportional to the equations of motion)  be rewritten in the form  $F\pa F \pa F \pa^2 F$. By using the Bianchi identity any $F \pa F \pa F \pa^2 F$ term can be expressed in terms of the following thirteen terms 
\bea
L_1 &=& F_{\mu\nu} \pa^{\la_1} F^{\eta\mu} \pa^{\la_2} F^{\de\nu} \pa_{\la_1}\pa_{\la_2} F_{\eta\de}\,, \non \\
L_2 &=& F_{\mu\nu} \pa^{\eta} F^{\la_1\mu} \pa^{\la_2} F^{\de\nu} \pa_{\la_1}\pa_{\la_2} F_{\eta\de}\,, \non \\
L_3 &=& F_{\mu\nu} \pa^{\eta} F^{\la_1\mu} \pa^{\de} F^{\la_2\nu} \pa_{\la_1}\pa_{\la_2} F_{\eta\de}\,, \non \\
L_4 &=& F_{\mu\nu} \pa^{\eta} F^{\de\mu} \pa^{\la_1} F^{\la_2\nu} \pa_{\la_1}\pa_{\la_2} F_{\eta\de}\,, \non \\
L_5 &=& F_{\mu\nu} \pa_{\rho} F^{\la_1\mu} \pa^{\eta} F^{\la_2\rho} \pa_{\la_1}\pa_{\la_2} F_{\eta}{}^{\nu}\,, \non \\
L_6 &=& F_{\mu\nu} \pa^{\la_1} F_{\rho}{}^{\mu} \pa^{\la_2} F^{\eta\rho} \pa_{\la_1}\pa_{\la_2} F_{\eta}{}^{\nu}\,, \non \\
L_7 &=& F_{\mu\nu} \pa^{\la_1} F_{\rho}{}^{\mu} \pa^{\eta} F^{\la_2\rho} \pa_{\la_1}\pa_{\la_2} F_{\eta}{}^{\nu}\,, \non \\
L_8 &=& F_{\mu\nu} \pa_{\rho} F^{\eta\mu} \pa^{\la_1} F^{\la_2\rho} \pa_{\la_1}\pa_{\la_2} F_{\eta}{}^{\nu} \non\,, \\
L_9 &=& F_{\mu\nu} \pa^{\eta} F_{\rho}{}^{\mu} \pa^{\la_1} F^{\la_2\rho} \pa_{\la_1}\pa_{\la_2} F_{\eta}{}^{\nu}\,, \non \\
L_{10} &=& F_{\mu\nu} \pa_{\rho} F^{\la_1\mu} \pa^{\la_2} F^{\eta\rho} \pa_{\la_1}\pa_{\la_2} F_{\eta}{}^{\nu} \,,\non \\
L_{11} &=& F_{\mu\nu} \pa_{\rho} F_{\eta}{}^{\la_1} \pa^{\rho} F^{\eta\la_2} \pa_{\la_1}\pa_{\la_2} F^{\mu\nu} \,,\non \\
L_{12} &=& F_{\mu\nu} \pa_{\rho} F_{\eta}{}^{\la_1} \pa^{\eta} F^{\rho\la_2} \pa_{\la_1}\pa_{\la_2} F^{\mu\nu} \,,\non \\
L_{13} &=& F_{\mu\nu} \pa^{\rho} F^{\mu\nu} \pa^{\la_1} F^{\la_2\eta} \pa_{\la_1}\pa_{\la_2} F_{\eta\rho} \,.
\eea
By integrations by parts there will be relations between these
terms. There are nine independent such relations, reducing the number
of independent elements to four. These can be chosen to be $L_1$,
$L_6$, $L_{11} - L_{12}$ and $L_{13}$. By integrations by parts and
use of the Bianchi identity these four quantities can in turn (modulo
the equation of motion) be expressed in terms of the following ones
\bea
X_1 &=& \pa_{\rho} F_{\mu\nu} \pa^{\rho}F^{\nu\mu}  \pa_{\la} F_{\eta\de} \pa^{\la}F^{\de\eta} = \tr(\pa_{\rho}F\pa^{\rho}F) \tr(\pa_{\la}F\pa^{\la}F) \,,\non \\
X_2 &=&   \pa_{\rho} F_{\mu\nu} \pa_{\la}F^{\nu\mu}  \pa^{\rho}
F_{\eta\de} \pa^{\la}F^{\de\eta} = \tr(\pa_{\rho}F\pa_{\la}F)
\tr(\pa^{\rho}F\pa^{\la}F) \,, \non \\ 
X_3 &=& \pa_{\rho}F_{\mu\nu}\pa^{\rho}F^{\nu\eta}\pa_{\la}F_{\eta\de}\pa^{\la}F^{\de\mu} = \tr (\pa_{\rho}F\pa^{\rho}F\pa_{\la}F\pa^{\la}F)\,,\non \\
X_4 &=& \pa_{\rho}F_{\mu\nu}\pa_{\la}F^{\nu\eta}\pa^{\rho}F_{\eta\de}\pa^{\la}F^{\de\mu} = \tr (\pa_{\rho}F\pa_{\la}F\pa^{\rho}F\pa^{\la}F)\,.
\eea
For completeness we list the relations between the $X$'s and the $L$'s
(valid up to terms proportional to the lowest order equation of motion) 
\be
\ba{rclrcl}
L_1 &=& \ts{\frac{1}{2}}X_4 - X_3 \,,& L_2 &=& \ts{\frac{1}{32}}(-X_1 + 6 X_2 -24 X_3 + 12 X_4)\,, \non \\
L_3 &=& \ts{\frac{1}{16}}(-X_1+6X_2 -8X_3+4X_4)\,, & L_4&=&\ts{\frac{1}{32}}(-X_1+6X_2+8X_3-4X_4)\,, \non \\
L_5 &=& 0\,, & L_6 &=& \ts{\frac{1}{2}}X_4\,, \non \\
L_7 &=& \ts{\frac{1}{32}}(-3 X_1 + 10 X_2 - 8 X_3 + 20 X_4)\,, & L_8 &=& 0\,, \non \\
L_9 &=& \ts{\frac{1}{32}}(-X_1 - 2X_2 + 8 X_3 + 12 X_4) \,, & L_{10} &=& \ts{\frac{1}{32}}(X_1 + 2 X_2 - 8X_3 + 4 X_4)\,, \non \\ 
L_{11} &=& \ts{\frac{1}{8}}(X_1+2X_2-8X_3 + 4 X_4)\,, & L_{12} &=&\ts{\frac{1}{8}}(-X_1 + 6X_2 - 8X_3 + 4X_4)\,, \non \\
L_{13} &=&\ts{\frac{1}{4}}X_1\,.
\ea
\ee
We now have all information needed to check that (\ref{BI2}) indeed contains the
result (\ref{t8F4}). Most of the contributions in (\ref{BI2}) can be
straightforwardly reexpressed in terms of the $X$'s, whereas some
require the use of the above results. We will not give any further
details here. 

The expression (\ref{t8F4}) can be rewritten in terms of
the $t_8$ tensor. When this eight-index tensor is contracted with four
arbitrary anti-symmetric matrices the result is 

\bea
t_{8}^{\mu_1\cdots
\mu_8}A_{\mu_1\mu_2}B_{\mu_3\mu_4}C_{\mu_5\mu_6}D_{\mu_7\mu_8} &:=&-\, 2[\tr(AB)\tr(CD)+\tr(AC)\tr(BD) +\tr(AD)\tr(BC)]
\non \\ && +\,  8\,\tr(ABCD + ACBD + ADCB)
\eea

\setcounter{equation}{0}
\section{Sums galore} \label{sumapp}
We here collect some of the sums which appear in the
 calculations. Some of the
 expressions involve the dilogarithm, $\Li_2(x)$, defined as $\Li_2(x)=\sum_{k=1}^{\infty} \frac{x^k}{k^2}$. Various special values 
 are $\Li_2(1) = \frac{\pi^2}{6}$ and
 $\Li_2({-1})=-\frac{\pi^2}{12}$. The dilogarithm has the following
 asymptotic expansion for large $x$: $\Li_2(-x) \sim -\half \ln^2(x) -
 \frac{\pi^2}{6}$. The following relations between $\Li_2({-\half})$,
 $\Li_{2}(-2)$ and $\Li_2(-3)$ are also useful: 
\bea
\Li_2(-\half)+ \Li_2(-2) &=&
 -\ts{\frac{\pi^2}{6}} \,, \non \\
 \Li_2(-2)+\half \Li_2(-3)
 &=& -\ts{\frac{\pi^2}{6}} - \ln2\ln3 \,.
\eea  
Further properties and relations can be found in \cite{Lewin}. We now
list the sums over integers which are needed in this paper.  
In the expressions below $a,b\ge 0$ and $\ep>0$. The notation $\sim$
should be read `equal up to $\cO(\ep)$ corrections'. 
\bea
\displaystyle{\sum_{n=1}^{\infty}} e^{-\ep n} &=& \frac{e^{-\ep} }{ 1-e^{-\ep} }
\sim \ts{\frac{1}{\ep}} - \half \non \\
\displaystyle{\sum_{n=1}^{\infty}} \mbox{$\frac{e^{-\ep n}}{n}$} &=&-\ln (1{-}e^{-\ep}) \sim
-\ln\ep \non \\
\displaystyle{\sum_{m,n=1}^{\infty}} \mbox{$\frac{e^{-\ep (m+n)}}{m{+}n}$} &=&\ln (1{-}e^{-\ep})
+\frac{e^{-\ep} }{ 1-e^{-\ep}} \sim  \ts{\frac{1}{\ep}} + \ln\ep -
\half \non \\
\displaystyle{\sum_{m,n=1}^{\infty}}\mbox{$\frac{e^{-\ep (m+n)}}{m(m{+}n)}$}
&=& \half \ln^2(1{-}e^{-\ep}) \sim \half \ln^2 \ep \non \\
\displaystyle{\sum_{m,n=1}^{\infty}}\mbox{$\frac{e^{-\ep (m+n)}}{(m{+}n)^2}$}
&=& -\ln (1{-}e^{-\ep}) - \Li_2(e^{-\ep})\sim -\ln \ep +
\mbox{$\frac{\pi^2}{6}$} \non \\
\ds{\sum_{n,m,p,q=1}^{\infty}}\ts{\frac{e^{-\ep(m+n+p+q)}}{m
 n}}\scriptstyle{\de(m{+}p{-}n{-}q) }&=&\half \ts{\frac{ e^{-2\ep} }{
 1-e^{-2\ep} }}\Big[\ln^2(1{-}e^{-2\ep}) - 2 \Li_2(-\ts{\frac{
 e^{-2\ep} }{ 1-e^{-2\ep} }})\Big] \non \\ & \sim&
 \left(\ts{\frac{1}{2\ep}-\frac{1}{2}}\right)[\ts{\frac{\pi^2}{6}}+\mbox{log terms}]\non \\
\displaystyle{\sum_{m,n=1}^{\infty}}\mbox{$\frac{e^{-\ep
(b(m+n){+}am)} }{m(m{+}n)}$}
&=& \ln (1{-}e^{-\ep a})\ln (e^{-\ep a}\mbox{$\frac{1-e^{-\ep
 b}}{1-e^{-\ep a}}$})-\Li_2(-e^{-\ep a}\mbox{$\frac{1-e^{-\ep b}}{1-e^{-\ep a}}$}) \non \\ &&+
 \Li_2(-\mbox{$\frac{e^{-\ep a}}{1-e^{-\ep a}}$}) - \ln (1{-}e^{-\ep a})\ln (\mbox{$\frac{e^{-\ep
 a}}{1-e^{-\ep a}}$}) \non \\ &\sim& -\Li_2(-\ts{\frac{b}{a}}) -
\ts{\frac{\pi^2}{6}} + \mbox{log terms}\non \\
\ds{\sum_{n,m}^{\infty}} \ts{\frac{e^{-\ep(b(m+n) + a n})}{m(m{+}n)}}
&=& \ln(1{-}e^{-\ep b
})\ln[\ts{\frac{1{-}e^{-\ep(b{+}a)}}{1-e^{-\ep a}}}] + \Li_2(-e^{- \ep
a } \ts{\frac{1-e^{-\ep b}}{1-e^{-\ep a}}}) \non \\ &&-
\Li_2(-\ts{\frac{e^{- \ep a}}{1-e^{-\ep a}}}) \non \\ &\sim&\Li_2(-\ts{\frac{b}{a}}) +
\ts{\frac{\pi^2}{6}} + \mbox{log terms} \non \\
\ds{\sum_{m,n,k=1}^{\infty}} \ts{ \frac{ e^{-\ep(b(m+n+k) +
a(m+k))} }{ m(m+n+k) } } &=&
\ts{\frac{ e^{-\ep a} }{ 1-e^{-\ep a} }
}\Big[\ln(1{-}e^{-\ep(a+b)})\ln(1{-}e^{-\ep a}) - \ts{\frac{e^{\ep
a}}{2}} \ln^2 (1{-}e^{-\ep(a+b)})  \non \\ && - \ln (1{-}e^{-\ep b})
\ln(\ts{\frac{1{-}e^{-\ep (b{+}a)}}{1-e^{-\ep a}}}) -   \Li_2(-e^{-\ep
a}\ts{\frac{1{-}e^{-\ep b}}{1-e^{-\ep a}}}) \non \\ &&+
\Li_2(-\ts{\frac{e^{-\ep a}}{1-e^{-\ep a}}}) \Big]  \non    \\
&\sim&-\left(\ts{\frac{1}{a\ep}
-\frac{1}{2}}\right)[\Li_2(-\ts{\frac{b}{a}}) + \ts{\frac{\pi^2}{6}}]
+ \mbox{log terms}
\non \\
\ds{\sum_{m,n,k=1}^{\infty}} \ts{ \frac{ e^{-\ep(b(m+n+k) +
ak)} }{ m(m+n+k) } } &=& \half
\ts{\frac{ 1 }{ 1-e^{-a\ep} }
}\Big\{\ln( 1{-}e^{-\ep b} )[e^{-\ep a} \ln( 1{-}e^{-\ep b} ) -2\ln( \ts{\frac{1{-}e^{-\ep
(a+b)} } { 1{-}e^{-\ep a} } }) ]  \non \\ 
&& - 2\Li_2(-e^{-\ep
a}\ts{\frac{1{-}e^{-\ep b}}{1-e^{-\ep a}}})
+2\Li_2(-\ts{\frac{e^{-\ep a}}{1-e^{-\ep a}}}) \Big\}  \non \\ &\sim&-
\left(\ts{\frac{1}{a\ep} +\frac{1}{2}}\right)[\Li_2(-\ts{\frac{b}{a}})
+ \ts{\frac{\pi^2}{6}}] + \mbox{log terms} \non \\
\ds{\sum_{m,n,k=1}^{\infty}} \ts{ \frac{ e^{-\ep(b(m+n+k) +
a m )} }{ (m+n)(m+n+k) } } &=& \ts{\frac{1 }{ 1-e^{-a\ep} }
}\Big[-\ln(1{-}e^{-\ep(a+b)})\ln(1{-}e^{-\ep a}) + \ts{\frac{e^{-\ep
a}}{2}} \ln^2 (1{-}e^{-\ep a})  \non \\ && + \ln (1{-}e^{-\ep b})
\ln(\ts{\frac{1{-}e^{-\ep (b{+}a)}}{1-e^{-\ep a}}}) +   \Li_2(-e^{-\ep
a}\ts{\frac{1{-}e^{-\ep b}}{1-e^{-\ep a}}}) \non \\ &&-
\Li_2(-\ts{\frac{e^{-\ep a}}{1-e^{-\ep a}}}) \Big]  \non \\ &\sim&\left(\ts{\frac{1}{a\ep} +\frac{1}{2}}\right)[\Li_2(-\ts{\frac{b}{a}})
+ \ts{\frac{\pi^2}{6}}] + \mbox{log terms} \non \\ 
\ds{\sum_{m,n,k=1}^{\infty}} \ts{ \frac{ e^{-\ep(4m+2k+2n)} }{
(m+n)(m+k) } } &=& \ts{\frac{2}{ 1-e^{-\ep} } }\ln(1{+}e^{-\ep}) -
\ln^2(1{-}e^{-\ep}) + \ln(1{-}e^{-\ep}) \non \\
&&- \ln(1{+}e^{-\ep}) \sim \mbox{log terms}\,. 
\eea
The method we used to derive these expressions was by first differentiating
with respect to either $\ep$, $a$ or $b$ in such a way as to obtain a simpler sum. Integration then gives the original sum. The integration constant can be fixed using the fact that the sum goes to zero when the parameter goes to infinity.
The sums over half-integer modes needed in this paper can be simply
related to the above sums by replacing all half-integer modes by $n{-}\half$, with $n\in \Z_+$, and possibly also shifting some of the
summation variables. Examples include 
\bea
\ds{\sum_{r=\frac{1}{2}}^{\infty}} e^{-\ep  r} &=& e^{\frac{\ep}{2} } \ds{\sum_{n=1}^{\infty}} e^{-\ep n} \,,\non \\
\ds{\sum_{r,s=\frac{1}{2}; k=1}^{\infty}} \ts{ \frac{ e^{-\ep(b(r+s+k) +
a r)} }{ (r+s)(r+s+k) } } &=& e^{\frac{\ep a}{2}}\Big[\ds{\sum_{m,n,k=1}^{\infty}} \ts{ \frac{ e^{-\ep(b(m+n+k) +
a m)} }{ (m+n)(m+n+k) } } + \ds{\sum_{m,k=1}^{\infty}} \ts{ \frac{ e^{-\ep(b(m+k) +
a m)} }{ m(m+k) } } \Big]\,.
\eea
We also need the following sums whose complete expressions
are rather involved; we therefore only list the relevant terms
\bea
\ds{\sum_{n,m,k=1}^{\infty}} \ts{ \frac{ e^{-\ep(4 m + 3n + 3k)} }{
(m+k)(m+n) } } - \ds{\sum_{r,s,t=\frac{1}{2}}^{\infty}} \ts{ \frac{ e^{-\ep(4 r + 3s + 3t)} }{
(r+s)(r+t) } } &\sim& \Li_2(-3) + \half \ts{\frac{\pi^2}{6}} +
\mbox{log terms}  \,,\non \\
\ds{\sum_{n,m,k=1}^{\infty}} \ts{ \frac{ e^{-\ep(3 m + 2n + 2k)} }{
(m+k)(m+n) } } - \ds{\sum_{r,s,t=\frac{1}{2}}^{\infty}} \ts{ \frac{ e^{-\ep(4 r + 3s + 3t)} }{
(r+s)(r+t) } } &\sim& \Li_2(-2) + \half \ts{\frac{\pi^2}{6}} +
\mbox{log terms}\,.  
\eea

\begingroup\raggedright\endgroup

\end{document}